\documentclass[aps,twocolumn,nofootinbib,floats,unsortedaddress,prd]{revtex4}
\usepackage{graphicx, epsfig, bm}
\def\mnras{Mon.~Not.~R.~Astron.~Soc.~}
\def\apjs{Astrophys.~J.~Suppl.~}
\usepackage{color}

\usepackage{hyperref}
\usepackage{ifthen}

\begin{document}

%%%%%%%%%%%%%%%%%%%%%%%%%%%%%%%%%%
\newcommand{\zmin}{z_{\rm min}}
\newcommand{\zmax}{z_{\rm max}}
\newcommand{\dz}{\Delta z}
\newcommand{\dzsub}{\Delta z_{\rm sub}}
\newcommand{\zminsn}{z_{\rm min}^{\rm SN}}
\newcommand{\nzpc}{N_{z,{\rm PC}}}
\newcommand{\zinit}{z_{\rm init}}
\newcommand{\amax}{a_{\rm max}}
\newcommand{\atr}{a_{\rm tr}}
\newcommand{\aeq}{a_{\rm eq}}
\newcommand{\wmin}{w_{\rm min}}
\newcommand{\wmax}{w_{\rm max}}
\newcommand{\wfid}{w_{\rm fid}}
\newcommand{\lcdm}{$\Lambda$CDM}
\newcommand{\om}{\Omega_{\rm m}}
\newcommand{\orad}{\Omega_{\rm r}}
\newcommand{\ode}{\Omega_{\rm DE}}
\newcommand{\ok}{\Omega_{\rm K}}
\newcommand{\omhh}{\Omega_{\rm m} h^2}
\newcommand{\winf}{w_{\infty}}
\newcommand{\scrm}{\mathcal{M}}
\newcommand{\osf}{\Omega_{\rm sf}}
\newcommand{\omf}{\Omega_{\rm m}^{\rm fid}}
\newcommand{\scrmf}{\mathcal{M}^{\rm fid}}
\newcommand{\rhode}{\rho_{\rm DE}}
\newcommand{\rhoc}{\rho_{{\rm cr},0}}
\newcommand{\dlum}{d_{\rm L}}
\newcommand{\dlss}{D_*}
\newcommand{\zmaxpc}{\zmax^{\rm PC}}
\newcommand{\zmaxlike}{\zmax^{\rm \mathcal{L}}}
\newcommand{\gpr}{G^{\prime}}
%%%%%%%%%%%%%%%%%%%%%%%%%%%%%%%%%%%%%%%%%%%

\pagestyle{plain}

\title{Falsifying Paradigms for Cosmic Acceleration}

\author{Michael J.\ Mortonson}
\affiliation{Kavli Institute for Cosmological Physics 
       and Department of Physics, Enrico Fermi Institute,
        University of Chicago, Chicago, IL 60637}

\author{Wayne Hu}
\affiliation{Kavli Institute for Cosmological Physics 
       and Department of Astronomy \& Astrophysics, Enrico Fermi Institute,
        University of Chicago, Chicago, IL 60637}

\author{Dragan Huterer}
\affiliation{Department of Physics, University of Michigan, 
450 Church St, Ann Arbor, MI 48109-1040}

\begin{abstract}
Consistency relations between growth of structure and expansion history
observables exist for any physical explanation of cosmic acceleration, be it
a cosmological constant, scalar field quintessence, or a general component
of dark energy that is smooth relative to dark matter on small scales.  The
high-quality supernova sample anticipated from an experiment like SNAP and
CMB data expected from Planck thus make strong predictions for growth and
expansion observables that additional observations can test and potentially
falsify.  We perform an MCMC likelihood exploration of the strength of these
consistency relations based on a complete parametrization of dark energy
behavior by principal components.  For \lcdm, future SN and CMB data make
percent level predictions for growth and expansion observables. For
quintessence, many of the predictions are still at a level of a few percent
with most of the additional freedom coming from curvature and early dark
energy.  While such freedom is limited for quintessence where phantom 
equations of state are forbidden, it is larger in the smooth dark energy class.
Nevertheless, even in this general class 
predictions relating growth measurements at different redshifts
remain robust, although predictions for the instantaneous growth rate do not.
Finally, if observations falsify the whole smooth dark
energy class, new paradigms for cosmic acceleration such as modified gravity
or interacting dark matter and dark energy would be required.
\end{abstract}
\maketitle

%\keywords{cosmology, large-scale structure of the  universe}

%=================================================================
\section{Introduction}
\label{sec:introduction}

A decade after the first firm evidence for the accelerated expansion of the
universe \cite{Riess_1998, Perlmutter_1999}, the study of dark energy remains
one of the most important yet difficult endeavors in theoretical cosmology
(e.g.\ \cite{Copeland_review,Linder_review,FriTurHut}).  The quality of data
from a variety of cosmological probes has strengthened in recent years
\cite{Garnavich_1998,Spergel_2003,Tonry_2003,Knop_2003,Tegmark_SDSS,Riess_2004,Seljak_SDSS,Eisenstein,Jarvis,Sanchez_2dF_2005,Astier,Spergel_2006,Tegmark_LRG,Riess_2006,WoodVasey_2007,Davetal07,PercivalBAO,Guzzo,Komatsu_2008,SCP_Union,Gaztanaga}, 
leading to multiple, independent lines of
evidence for the accelerating expansion. In the near future, we can expect a
battery of measurements with unprecedented precision that will provide
stringent tests of any purported explanation of cosmic acceleration.

Despite the tremendous amount of raw information expected from upcoming Type Ia
supernova (SN) surveys, baryon acoustic oscillations (BAO) from galaxy 
redshift surveys, weak lensing, and
cluster counting surveys, only a handful of parameters associated
with the dark energy equation of state can be constrained \cite{Linder_Huterer_howmany,Suletal07}.
This limitation arises because most observables depend on 
cosmological distances and growth, which are integrals over the expansion rate, 
which itself contains an integral over the dark energy equation of state.

Although insensitivity to fine-scale features of the equation of state is a
drawback for measuring dark energy parameters, it is an advantage for testing
the consistency among acceleration observables required by dark energy
paradigms; since the individual probes of dark energy do not depend strongly
on the rapidly oscillating evolution of the equation of state, neither do the
consistency relations between these observables.  For example, it is well
known that under a cosmological constant explanation of acceleration or simple
parametrizations of the equation of state, distance measurements predict the
growth of structure in a spatially flat universe.  Violation of this
consistency relation would falsify the standard flat \lcdm\ model and its most
basic generalizations.
 
The goal of this study is to extend these ideas of prediction and
falsification from simple dark energy parametrizations to general classes of
dark energy models with time-dependent equations of state, specifically scalar
field quintessence and dark energy that is spatially smooth compared with the
dark matter on small scales.  We start with SN and cosmic microwave background
(CMB) measurements expected in the next decade and make predictions for growth
and expansion history measurements as a function of redshift within dark
energy model classes parametrized by a complete basis of principal components.
Where predictions are tight, observations can falsify the model class.  Where
predictions are loose, observations can better pin down the parameters of the
class, in particular those controlling spatial curvature and dark energy that
is significant at early times.

Our study complements previous work
\cite{Huterer_Peiris,Chongchitnan_Efstathiou,Barnard_discrim} on the
observable predictions of classes of dark energy models, and follows a long
history of studies concerning the best way to probe dark energy using
cosmological observations
\cite{TegEisHuKron,EHT,Cooray_Huterer,Hu99,Huterer_Turner,Weller_Albrecht,Huterer_thesis,Kujat,Maor01,
  CorCop03,FriHutLinTur,HuHai03,Seo03,Hu_Jain,Huterer_Cooray,Feng,Wiley04,Hu_Scranton,Upadhye,Wang_Tegmark_2005,Glazebrook_Blake,Pogosian_ISW,Sahlen05,Crittenden_Pogosian,
  Xia,Hui_Greene,Zhao,Knox_Song_Zhan,DETF,Liddle_evidence,Sahlen07,Zhan_Knox_06,Crit_Maj_Piazza,Wang_Mukherjee_07,
  Barnard_AS,Abrahamse_pngb,Bozek_exp,ZhaKnoTys08,Rubin_SCP}.
To our knowledge this is the first time that consistency between growth,
distance, and other expansion observables has been studied in such a general
and quantitative way. In particular, while some previous studies have
considered predictions of dark energy described with a small number of
parameters (e.g.~\cite{Huterer_Peiris,Barnard_discrim,Sahlen07}), here we
consider $\sim 500$ parameters of which about $10-15$ are necessary to
completely describe to high accuracy the predictions of current or future
data.  Although we compute equations of state for dark energy as an
intermediate step between distance and growth, we emphasize that our goal is
\emph{not} to reconstruct $w(z)$, unlike many previous studies (e.g.\
\cite{reconstr,Starobinsky,Nakamura_Chiba,Saini_reconstr,Weller_Albrecht,
  Gerke_Efstathiou, Wang_Mukherjee_03,Sahni_review,Li_Holz_Cooray,
  Zunckel_Trotta}).  Rather than addressing the quality of
constraints on the equation of state, here we are more interested in using
dark energy parameters only as a tool to study how to falsify basic paradigms
for cosmic acceleration.

This paper is organized as follows. In \S~\ref{sec:prelim}, we describe our
methods for computing predictions for observables from future cosmological
data under different dark energy paradigms.  These predictions can lead to
falsification and subsequent generalization of each model class as shown in
\S~\ref{sec:dectree}.  The main tests of each class are summarized in
\S~\ref{sec:conc}.  Appendices provide additional details about the inclusion
of various data in the construction of principal components and in the
likelihood analysis, our methodology for computing the principal components,
and tests of the completeness of our parametrization in the growth and
expansion observables.

%=================================================================
\section{Methodology}
\label{sec:prelim}

%--------------------------------------
\subsection{Dark energy principal components}
\label{sec:pcs}

We parametrize the dark energy equation of state, $w(z)$, with a basis of 
principal components (PCs) \cite{Huterer_Starkman,Hu_PC}.  
The PC amplitudes are weighted redshift 
averages of $w(z)$ ordered by how well they are measured, and can
be straightforwardly computed for a given data set.  As we discuss in
\S~\ref{sec:observ} and~\ref{sec:tests}, principal components based on SN
distance modulus data in particular provide a nearly complete basis for other
acceleration observables.  We therefore treat the PC amplitudes as simply a
convenient intermediate representation to capture the information in the
distance modulus and translate it into predictions for other observables.

 Principal components can also be defined for other
redshift-dependent quantities such as
the dark energy density $\rhode(z)$ (e.g.\ \cite{Hu_PC,Dick})
which is related to $w(z)$ by
\begin{equation}
\rhode(z) = \rhoc \ode \exp\left[3 \int_0^z dz' \frac{1+w(z')}{1+z'}\right],
\end{equation}
where $\rhoc$ is the critical density and $\ode$ is the fraction of dark
energy, both at the present time.  Refs.~\cite{Linder04,WangFreese} discuss
the advantages of using either $w(z)$ or $\rhode(z)$ to describe dark energy.
Our choice to use $w(z)$ is motivated by the fact that the model classes we
consider are separated by the allowed values of $w$: $w=-1$ for \lcdm\ and
$-1\leq w\leq 1$ for quintessence. (Values of $w$ outside this range are
possible for quintessence models in which the dark energy density becomes
negative, but such models are inconsistent with current data as we discuss in
\S~\ref{sec:tree2}.)

Specifically, we compute the PCs based on distances to Type Ia supernovae with
measurement errors modeled after the proposed specifications for the
SuperNova/Acceleration Probe (SNAP \cite{SNAP}) experiment. We also assume
that 300 low-$z$ SNe will be available for calibrating the normalization of
the high-$z$ distance-redshift relation.  We supplement the SN observables
with constraints on the expansion history at earlier times from the CMB
acoustic peaks using the precision expected from the Planck mission. We take
the CMB observables to be the matter density scaled to the present, $\omhh$,
and the comoving angular diameter distance to last scattering, $\dlss \equiv
D(z_*)$ where $z_* \approx 1090$ \cite{Komatsu_2008}.  Our assumptions about
these fiducial SN and CMB data sets and additional priors are detailed in
Appendix~\ref{app:data}.  The fiducial cosmology we assume for the PC
construction (and for likelihood analysis) is flat \lcdm\ with present matter
fraction $\om=0.24$ and Hubble constant $h=0.73$, consistent with current data
\cite{Komatsu_2008,SCP_Union}.

\begin{figure}[t]
\centerline{\psfig{file=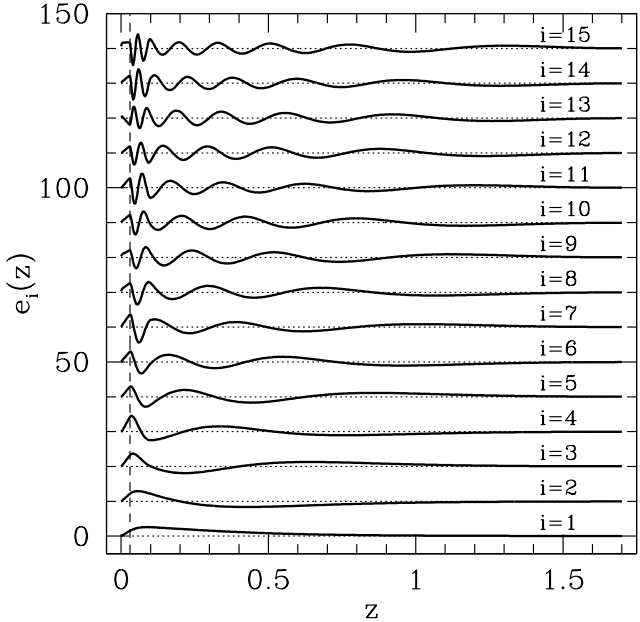, width=3.5in}}
\caption{The first 15 PCs of $w(z)$ (increasing variance from
  bottom to top), with 500 redshift bins between $z=0$ and
  $\zmax=1.7$. The vertical dashed line shows the minimum redshift of the data
  assumed for computing the PCs, $\zminsn = 0.03$; SNe are distributed between
  $\zminsn$ and $\zmax$ following a SNAP-like distribution (given in
  \cite{KLMM_SNAP}) plus a low-z sample at $\zminsn\leq z\leq 0.1$
  (see Appendix~\ref{app:data}).
  When computing the PCs we include a CMB prior modeled after Planck and
  marginalize over $\om$, $\omhh$, and the absolute magnitude of the SNe.  
  The PCs are offset vertically
  from each other for clarity with dotted lines showing the zero point for each
  component.  }
\vskip 0.25cm
\label{fig:pcs}
\end{figure}

The principal component functions $e_{i}(z_{j})$ are eigenvectors of the
covariance matrix for the equation of state in redshift bins $\{z_j\}$, 
and they form a basis in which an arbitrary
function $w(z_j)$ may be expressed as
\begin{equation}
w(z_j) - \wfid(z_j) = \sum_{i=1}^{\nzpc} \alpha_i e_i(z_j),
\label{eq:pcstow}
\end{equation}
where $\alpha_i$ are the PC amplitudes, $\nzpc = 1+\zmax/\dz$ is the number of
bins in redshift, and $z_j = (j-1) \dz$.  We adopt a maximum redshift for
variations in $w(z)$ of $\zmax=1.7$ to match the largest redshift for our
fiducial supernova data, and we use a fiducial model $\wfid(z)=-1$ since
\lcdm\ is an excellent fit to current data.  For these choices of $\zmax$ and
$\wfid(z)$, Fig.~\ref{fig:pcs} shows the 15 lowest-variance PCs, which form
the basis we use for likelihood analysis.  We comment on how the PCs depend on
the choices of $\zmax$ and the fiducial cosmology in Appendix~\ref{app:pcs}.

By normalizing the PCs as
\begin{equation}
\sum_{i=1}^{\nzpc} [e_i(z_j)]^2 = \sum_{j=1}^{\nzpc} [e_i(z_j)]^2 = \nzpc,
\label{eq:norm}
\end{equation}
the components approach a ``continuum limit'' as $\dz \to 0$ in which the
shapes of all but the worst-determined PCs become smooth and independent of
$\dz$ (or $\nzpc$).  A small bin width $\dz \lesssim \zminsn$ also allows us
to resolve changes in $w(z)$ at redshifts below that of the nearest supernova
in the sample, $\zminsn=0.03$, which can evade SN constraints as discussed in
Appendix~\ref{app:pcs}.  We have chosen to use bins spaced linearly in
redshift, and the exact shapes of the components depend somewhat on this
choice; for example, had we chosen bins with equal widths in $a=(1+z)^{-1}$ or
$\ln(1+z)$, the weights of the PCs would have shifted in redshift
\cite{dePutter_Linder}. However, the most important property of the PCs for
our purposes is that they form a complete basis for observables such as
distance and growth (\S~\ref{sec:tests}), and this completeness can be achieved
for a variety of different binning conventions.

Although we use a large number of redshift bins ($\nzpc\sim 500$) to approach
the continuum limit of the PC shapes, we generally truncate the sum in
Eq.~(\ref{eq:pcstow}) to include only the $N_c < \nzpc$ modes that are
measured best by the fiducial data.  Predictions in \S~\ref{sec:dectree} are
based on a choice of $N_c=15$, and we explain how this number of PCs ensures
completeness in various observables in \S~\ref{sec:tests}.

% ****************************************
% ****************************************

For model classes that restrict $w(z)$ to some range $\wmin \leq w \leq
\wmax$, we can place priors on the PC amplitudes analogous to those introduced
in Ref.~\cite{MorHu08} for reionization principal components.  These priors,
which we define in Appendix~\ref{app:data}, include top-hat bounds on each PC
amplitude and an upper limit on the sum of squares of the amplitudes.  Due to
the truncation of the number of principal components required for likelihood
analysis, we adopt conservative priors.  A combination of PC amplitudes is
only excluded if the resulting equation of state at some redshift exceeds the
bounds on $w$ {\it regardless} of the amplitudes of the truncated components
($\{\alpha_i\}$ with $i>N_c$).  Conversely, satisfying the PC priors does not
guarantee that a reasonable set of truncated components can bring $w(z)$ back
within the bounds.  The priors therefore include all models within a class,
but do not necessarily exclude all models outside that class.

Our baseline dark energy model class is parametrized by the PC amplitudes, $\om$, 
and $\omhh$ in a flat universe:
 \begin{equation}
\bm{\theta}_{\rm base}=\{\alpha_1,\ldots, \alpha_{N_c}, \om, \omhh\}\,.
\label{eq:parametersbase}
\end{equation}
The Hubble constant, $h=H_0/(100~{\rm km~s}^{-1}{\rm Mpc}^{-1}) =
(\omhh/\om)^{1/2}$ is a derived parameter in this representation.

In this baseline class we take the dark energy density to be constant at $z>\zmax$.
The underlying assumption in this class is that by $z=\zmax$ the dark energy is already
much smaller than the matter density as in \lcdm,
and enforcing constant dark energy density at $z>\zmax$ assures that
it becomes increasingly irrelevant at higher redshift.
Note that our baseline model class includes the standard \lcdm\ model
of a cosmological constant in a flat universe, corresponding to 
$\{\alpha_i\}=0$ for $\wfid(z)=-1$.

\subsection{Early Dark Energy and Curvature}

If observations falsify the baseline model class, we can generalize it by
including both dark energy that remains a substantial fraction of the energy
density at $z>\zmax$, dubbed ``early dark energy'' 
\cite{Doran_Schwindt_Wett,Doran_Robbers}, and spatial curvature.

To describe early dark energy, we adopt a simple parametrization by 
assuming a constant equation of state, $w(z>\zmax)=\winf$
\cite{dePutter_Linder}.  The dark energy density at $z>\zmax$ can be
extrapolated from its value at $\zmax$ as
\begin{equation}
\rhode(z) = \rhode(\zmax)\left(\frac{1+z}{1+\zmax}\right)^{3(1+\winf)}.
\end{equation}
This description notably accounts for, but is not limited to, scalar field 
models that ``track" at $z>z_{\rm max}$
\cite{Ratra_Peebles,Ferreira_Joyce,Steinhardt_tracking} where the equation of
state is determined by that of the dominant component, in this case matter ($w=0$).
We examine the limitations of this parametrization in Appendix~\ref{app:complete}.
Instead of $\winf$, we use $\exp(\winf)$ as the parameter for likelihood analysis since 
models with $\winf \ll -1$ all have rapidly vanishing dark energy density at 
$z>\zmax$ and are therefore degenerate with each other in all 
observables. We allow the early dark energy parameter to vary within 
the range $0 \leq \exp(\winf) \leq 1$, where the upper limit eliminates $\winf>0$ models 
with dark energy density that exceeds the matter density at early times.
We restrict the allowed range for $\winf$ further in model classes where 
the low redshift equation of state is bounded (see Appendix~\ref{app:data}
for details).

Note that components of dark matter that are smooth
on small scales, for example very light neutrinos, 
are also described by the early dark energy parametrization. We will not distinguish
between these two possibilities here as that would require measurements in 
a regime where either the neutrinos or the dark energy were not smooth.

To complete our most general model class, we allow for the possibility of spatial curvature,
parametrized by $\ok \equiv 1-\om-\ode$.  The full parametrization for a 
dark energy model class is therefore
\begin{equation}
\bm{\theta}_{\rm full}=\{\alpha_1,\ldots, \alpha_{N_c}, \om, \omhh, \exp(\winf), \ok \}.
\label{eq:parameters}
\end{equation}
The  present dark energy density $\ode$ is derived from 
this parameter set. Setting $\winf=-1$ and $\ok=0$ recovers the 
baseline model class of Eq.~(\ref{eq:parametersbase}).

%--------------------------------------
\subsection{Markov Chain Monte Carlo}
\label{sec:mcmc}

We use the Markov Chain Monte Carlo (MCMC) algorithm to estimate 
the joint posterior distribution of cosmological parameters and 
derived observables by sampling the parameter space 
and evaluating the likelihood of each proposed model compared with an 
assumed data set 
(e.g.\ see~\cite{Christensen:2001gj,Kosowsky:2002zt,Dunetal05}).  
The posterior distribution is obtained using Bayes' Theorem,
\begin{equation}
{\cal P}(\bm{\theta}|{\bf x})=
\frac{{\cal L}({\bf x}|\bm{\theta}){\cal P}(\bm{\theta})}{\int d\bm{\theta}~
{\cal L}({\bf x}|\bm{\theta}){\cal P}(\bm{\theta})},
\label{eq:bayes}
\end{equation}
\noindent where ${\cal L}({\bf x}|\bm{\theta})$ is the likelihood of
the data ${\bf x}$ given the model parameters $\bm{\theta}$ and ${\cal
P}(\bm{\theta})$ is the prior probability density. The MCMC algorithm
generates random draws  from the posterior distribution that
are fair samples of the likelihood surface. From these samples, we can
estimate many properties of the posterior distribution including 
the mean values, covariance, and confidence intervals of both the 
basic set of parameters and derived parameters and observables.
Convergence of the set of random samples 
to a stationary distribution that approximates 
the joint posterior density ${\cal P}(\bm{\theta}|{\bf x})$ 
requires a large number of independent samples.
We use a minimum of four chains per model and determine when these chains 
have a sufficient number of samples for convergence by applying a conservative 
Gelman-Rubin criterion \cite{gelman/rubin} of $R-1\lesssim 0.01$.

The full details of the simulated cosmological data and priors used for the 
MCMC analysis and their likelihood functions are given in Appendix~\ref{app:data} and summarized
in \S \ref{sec:observ}.

We assume that all of the fiducial data are consistent with a flat 
\lcdm\ model with $\om=0.24$ and $h=0.73$, given that this model  
 fits current constraints well. We therefore do not consider here 
the potential for SN and CMB data to test this fiducial cosmology. 
It is, however, possible that these future measurements will 
falsify flat \lcdm\ by themselves, even before considering 
consistency with additional observables such as growth. 
We have checked that most of our qualitative conclusions do
not change with allowed alterations of the model underlying the 
SN and CMB data, and we note exceptions in Appendix~\ref{app:complete}.

Given a parametrization for a model class and the fiducial data, the 
MCMC posterior distribution then provides observable
predictions for parameters and derived acceleration observables 
that can be used as consistency tests to attempt to
falsify the whole model class.

%--------------------------------------
\subsection{Acceleration observables}
\label{sec:observ}

In this section, we define a set of redshift dependent observables that can be
probed by future experiments.  We focus on acceleration observables that can
be simply computed from the expansion history, leaving for future study the detailed relation
of these quantities to what is actually expected to be measured by specific
planned experiments.

We divide the observables into two categories.  In the first are observables
that we assume will be measured by a SNAP-like sample of supernovae and the
Planck satellite.  These measurements constitute the data for MCMC likelihood
analysis, from which we make predictions for the second category of
observables in specific model classes.

Supernova observations constrain the distance modulus, or relative luminosity
distance, between objects of different redshift in the sample.  We take the SN
data as a starting point since of the known methods for constraining the
acceleration, it has the finest resolution in redshift and hence its principal
components form the most complete set for providing testable predictions.
Supernovae have an additional advantage of being sensitive to low redshifts
$z\lesssim 0.5$ where dark energy dominates the energy budget and where other
probes like BAO and weak lensing do not have enough volume and distance,
respectively, in order to strongly constrain dark energy.  We take the same
Planck CMB constraints on $\omhh$ and angular diameter distance $\dlss$ as
used in the PC construction.

The SN and CMB data make predictions within a model class for 
the remaining observables, which include the expansion rate $H(z)$, 
the absolute distance $D(z)$, the growth function $G(z)$, and the growth
rate $f(z)$.   

The expansion rate, allowing for a general dark energy component and 
spatial curvature, is
\begin{equation}
H(z) = H_0 \left[ \om (1+z)^3 + \frac{\rhode(z)}{\rhoc} + \ok (1+z)^2 \right]^{1/2}.
\label{eq:hz}
\end{equation}
Except when dealing with CMB observables, we generally ignore the contribution of radiation 
to the expansion rate since it is a negligible fraction of the density at low $z$.
The absolute distance observable we use is the {\it comoving} (angular diameter) distance
\begin{equation}
D(z) = \frac{1}{(|\ok|H_0^2)^{1/2}} S_{\rm K}\left[(|\ok|H_0^2)^{1/2} 
\int_0^z \frac{dz'}{H(z')} \right],
\label{eq:dist}
\end{equation}
where the function $S_{\rm K}(x)$ is equal to $x$ in a flat universe ($\ok=0$), 
$\sinh x$ in an open universe ($\ok>0$), and $\sin x$ in a closed universe ($\ok<0$). 
Luminosity distances, whose ratios are measured by SNe (see Appendix~\ref{app:data}),
are simply related to Eq.~(\ref{eq:dist}) by $\dlum(z) = (1+z) D(z)$.

We define the growth function as 
$G(z) \propto (1+z)D_1(z)$, where $D_1(z)\equiv \delta(z)/\delta(\zinit)$ 
describes growth of the matter overdensity $\delta$ 
normalized at an initial redshift $\zinit=1000$ 
during matter domination. 
If $\om(z)\equiv \om (1+z)^{3} [H_0/H(z)]^2 =1$,
then $D_1(z)\propto (1+z)^{-1}$ and $G(z)$ is constant.  
The growth function obeys
\begin{equation}
G'' + \left(4+\frac{H'}{H}\right)G' + \left[
3+\frac{H'}{H}-\frac{3}{2}\om(z)\right]G = 0,
\label{eq:growth}
\end{equation}
where primes denote derivatives with respect to $\ln a$.
We normalize $G$ to its value at $\zinit$, taking the initial conditions to be
\begin{equation}
G(\zinit)=1,\quad G'(\zinit)= - \frac{3}{5}(1-\winf)\ode(\zinit),
\end{equation}
where $G'$ follows from a power-law solution to Eq.~(\ref{eq:growth}) 
assuming $\ode(\zinit)\ll 1$ and 
neglecting curvature since it has little effect on the expansion rate at early times.
We do not include radiation when solving Eq.~(\ref{eq:growth}), and 
we assume that the dark energy component is smooth on scales below the 
horizon at $z<\zinit$.

% ****************************************
\begin{figure}[t]
\centerline{\psfig{file=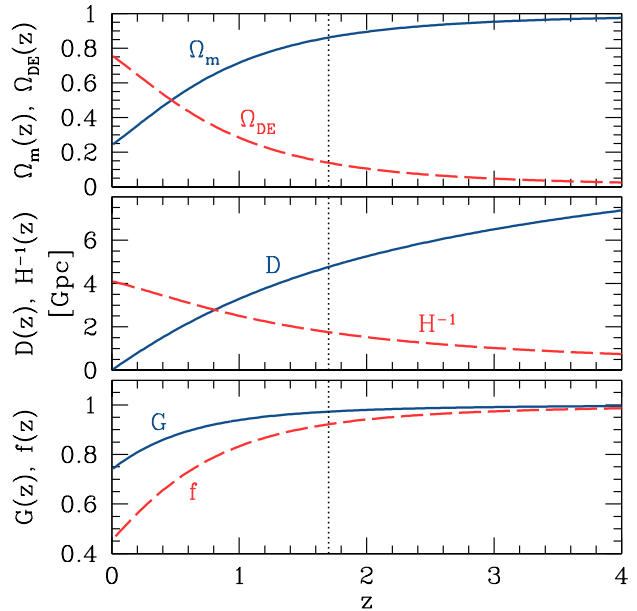, width=3.5in}}
\caption{Redshift dependent quantities for the fiducial 
flat \lcdm\ cosmology with $\om=0.24$ and $h=0.73$ 
as assumed for PC construction and for the default data sets for MCMC, 
including the fractions of the total density in matter and 
dark energy, $\om(z)$ 
and $\ode(z)$ (\emph{top; solid blue and dashed red, respectively}), 
comoving angular diameter distance 
$D(z)$ (\emph{middle, solid blue}), inverse of the expansion rate $H^{-1}(z)$ 
(\emph{middle, dashed red}), growth function relative to early times 
(\emph{bottom, solid blue}), and growth rate $f=1+d\ln G/d\ln a$ 
(\emph{bottom, dashed red}). The vertical dotted line is plotted 
at $\zmax=1.7$, the maximum redshift for the PCs and for SNe in 
the likelihood analysis.
  }
\vskip 0.25cm
\label{fig:fid}
\end{figure}
% ****************************************

Because growth measurements can be compared either to low redshift
data sets to obtain the relative growth $G_0(z)\equiv G(z)/G(z=0)$ between a 
redshift $z$ and the present, or to recombination through the CMB acoustic peaks
to obtain $G(z)$, 
we show predictions for both in the following sections.  The latter will ultimately
be limited by the measurement of the optical depth to reionization from CMB polarization
due to the translation between the observed acoustic peak amplitude and 
the intrinsic fluctuations at recombination.

The logarithmic growth {\it rate} is defined as
\begin{equation}
f(z) \equiv \frac{d\ln D_1}{d\ln a} = 1 + \frac{G'}{G},
\end{equation}
which is commonly approximated as $f(z) = \om^{\gamma}(z)$ 
where the growth index is $\gamma \approx 0.55$ for flat \lcdm\ 
\cite{WangSteinhardt,Linder_gamma}.
Measurements of $\gamma$ have been proposed as a way 
to test general relativity; we examine this idea in the 
context of various classes of cosmological models in \S~\ref{sec:tree4}.

In this paper, we remain agnostic about the techniques that best probe these
observables and simply assess the precision to which they can be predicted
in certain model classes.  However, some caveats are useful to keep in mind.
Although we make predictions as a function of redshift 
that can include fine-scale features, measurements will typically only
constrain coarse-grained averages of the predictions over wide bands in 
redshift.   For example, observations of the imprint of baryon acoustic oscillations 
(BAO) on galaxy clustering in directions transverse to the line of sight 
provide a standard ruler to constrain absolute distances $D(z)$, and 
BAO measurements along the line of sight can constrain 
the expansion rate, $H(z)$.  However, a volume of $\gtrsim 1$~Gpc$^3$ is required to obtain
accurate measurements of either quantity, resulting in smearing in
redshift.

Likewise, weak lensing measures both the growth function and ratios of
distances, but the broadness of the lensing kernel and scatter in photometric
redshifts again prevents a purely local measurement.   Growth rate
measurements that involve the redshift space distortion of galaxy surveys and
galaxy bias information from lensing suffer from broadening from both data sets.
Growth and growth rate measurements from the cluster abundance or weak lensing
in the nonlinear regime also
involve integrals over the past history of  growth and not  merely the instantaneous
linear growth \cite{Ma06}.

Finally, it is useful to place weak current priors on parameters related to the
observables.  In the broadest model classes that we consider, the SN 
and CMB measurements alone are not sufficiently predictive to eliminate even highly
deviant models.  To ensure that we only include models that are not already
ruled out by observations, we include priors on 
the fraction of dark energy at recombination $\ode(z_*)$ from WMAP
\cite{Doran}, the absolute distance 
$D(z=0.35)$ from current BAO measurements from SDSS \cite{Eisenstein}, 
and the Hubble constant $H_0$ from the HST Key Project \cite{HKP}.
The former represents the impact of the change in the growth function near
recombination on the first few acoustic peaks of the CMB.   We conservatively
do not include the expected improvement on this measurement from Planck.
 The BAO distance prior serves mainly 
to reduce the possible deviations from a flat geometry for dynamical dark 
energy models with spatial curvature.
The role of the Hubble constant 
prior in this analysis is to limit the variation in $w(z)$ at very low 
redshifts as we describe in Appendix~\ref{app:pcs}.

In summary, the observables that we predict  
are the expansion rate $H(z)$, comoving absolute distances $D(z)$, 
 the growth history $G(z)$ relative to recombination or $G_0\equiv G(z)/G(z=0)$ relative
 to the present, and the growth rate $f(z)$.
The redshift evolution of these quantities for the fiducial flat \lcdm\ model is plotted in 
Fig.~\ref{fig:fid}.
Predictions from SN and CMB data are derived by constructing these 
observables from models in the MCMC
samples described in \S~\ref{sec:mcmc},
and we explore the implications for various dark energy model classes in
\S~\ref{sec:dectree}.

%--------------------------------------
\subsection{Completeness}
\label{sec:tests}

To make predictions that can reliably be used to falsify paradigms for dark
energy, our parametrization must describe any effects that models within the
class might have on observables.  In particular, we must ensure that the set
of principal components that parametrize variation in $w(z)$ form a complete
basis for representing changes in growth and expansion observables relative to
the fiducial cosmology due to changes in the dark energy equation of state.
In this section we summarize our criteria for completeness, and refer the
reader to more detailed discussions of these issues in
Appendices~\ref{app:pcs} and~\ref{app:complete}.

When computing the PCs from the Fisher matrix for SN and CMB distances, we
have a choice of whether to fix or marginalize over the parameters other than
the binned $w(z)$. These decisions affect PC shapes due to degeneracies
between $w(z)$ and the other parameters.  We choose to fix $\ok$ and $\winf$,
thereby including degeneracies with curvature and early dark energy among the
well-measured PCs.  We marginalize $\om$ in the PC construction, thus reducing
the completeness for representing sharp transitions in $w(z)$ at $z<\zminsn$
as we describe in Appendix~\ref{app:pcs}.  Since these transitions are largely
indistinguishable and limited mainly by the external Hubble constant prior,
completeness is not important here.  We therefore choose in this instance to
sacrifice completeness for efficiency in representing the well constrained
redshift range.

Given the set of PCs, the next question is how many out of the full set of
$\nzpc$ components we need to keep as parameters (see \S~\ref{sec:pcs}). Note
that neglecting even the high variance PCs can have large effects on the
equation of state. However, since all of the observables contain integrals
over $w(z)$ the effects of these rapidly oscillating PCs (see
Fig.~\ref{fig:pcs}) tend to cancel out for the redshift dependent quantities
of interest, especially the distance and growth observables.  In general, the
number of components necessary for completeness, $N_c$, will be larger than
the number of dark energy parameters that can be measured to some specified
accuracy.  We typically find that the predicted range of observables changes
fractionally by less than a few percent between MCMC analyses with 10 and 15
PCs (see Appendix~\ref{app:complete}).  The agreement is somewhat worse for
$H(z)$ in some cases, but discrepancies occur mostly at $z<\zmax$ where
oscillations of $H(z)$ about the fiducial model would be averaged out in BAO
measurements over wide bins in redshift.  We conclude that $N_c \approx 10$ is
sufficient for completeness, but we present results from the larger set of 15
PCs to further reduce any remaining artifacts due to incompleteness.

The completeness of the parametrization could in principle depend
on the choice of fiducial model that we adopt to represent the true 
cosmology for future observations. We assume that the future SN and 
CMB data will be consistent with $\wfid=-1$ as is true of 
current measurements, and we examine an alternate choice of 
fiducial model in Appendix~\ref{app:complete}.

Completeness may also depend on the redshift range over which the PCs are
defined, $0\leq z\leq \zmax$.  The choice of $\zmax$ influences the definition
of early dark energy as well, since we ascribe any deviations from $w=-1$
behavior at $z>\zmax$ to early dark energy parametrized by constant $w=\winf$.
We find that choosing $\zmax=1.7$ to match the redshift coverage of the
fiducial SN data nicely balances between defining the PCs where cosmological
data have significant support (which argues for a lower $\zmax$) and having
the PCs be a complete representation for other observables (arguing for a
higher $\zmax$).  Additionally, our parameter $\winf$ is not intended to be a
complete description of early dark energy but rather a means of monitoring its
observable signatures. We present tests of both our choice of $\zmax$ and the
early dark energy parametrization in Appendix~\ref{app:complete}.

%=================================================================
\section{Testing dark energy paradigms}
\label{sec:dectree}

% ****************************************
\begin{figure}[t]
\centerline{\psfig{file=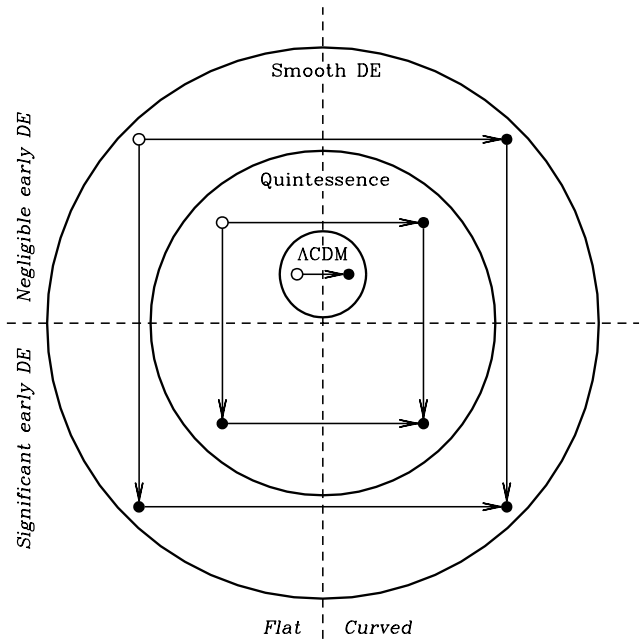, width=3.5in}}
\caption{Illustration of the model classes (large circles)
and subclasses (represented by points within different regions of the circles)
that we consider in this paper. Open points mark the initial (simplest) 
type of model within each class, and arrows indicate paths to the more 
complex models in the class. The \lcdm\ class does not contain 
models with significant early dark energy since the dark energy fraction 
vanishes rapidly at high $z$ for all allowed 
cosmological constant models.
}
\vskip 0.25cm
\label{fig:dectree}
\end{figure}
% ****************************************

Distance measurements from SNe and constraints from the CMB make predictions for the 
acceleration observables described in \S~\ref{sec:observ}  
that can be tested by future experiments. These predictions are made within
the context of a paradigm for acceleration, e.g.\ a cosmological constant.  Where the predictions are 
weak, the observables can be used to estimate parameters within 
the class, and where they are strong, precision measurements can potentially 
falsify the whole dark energy paradigm.

In the following sections, we step through the predictions for various dark
energy model classes.  Guided by Occam's razor and criteria for
falsifiability, we begin with the simplest model that satisfies current
constraints: flat \lcdm.  Models of this type make the firmest predictions and
are therefore easiest to falsify.  The next simplest and most predictive model
class is \lcdm\ with one additional parameter, spatial curvature.  This more
general class is particularly interesting in that falsification would rule out
a cosmological constant.

Since \lcdm\ corresponds to a constant dark energy equation of state $w=-1$,
our next step in generalizing the class of models is to allow $w$ to vary with
redshift.  We first consider a restricted range, $-1\leq w\leq 1$,
corresponding to the allowed values of the equation of state in quintessence
models where a canonical scalar field is responsible for cosmic acceleration
at late times (e.g.\ \cite{CalDavSte98}). 
Within the quintessence class, we study how the predictions
change when early dark energy and nonzero curvature are added to the basic
model.

Instead of going directly from a cosmological constant to arbitrary equations
of state, one could test several intermediate models along the way such as
constant $w$ or the two-parameter model $w(a)=w_0 + w_a(1-a)$
\cite{Chevallier_Polarski,Linder_wa}, and many authors have used this approach in
analyzing cosmological data.  Here, however, we take the view that unless the dark
energy falls into a well defined physical classification such as a cosmological
constant or a canonical scalar field, there is no reason to define a  
particular functional form of $w(z)$ as a class and not another.

For the final model class, we allow $w$ to vary over a much wider range than
in the case of quintessence but retain the requirement that dark energy is
smooth compared with dark matter on scales associated with the measurements of
growth.  For example, non-canonical kinetic terms can lead to equations of
state with $w < -1$ \cite{ArmMukSte00}.  If such a field has a sound speed
substantially below the speed of light, then the growth predictions presented
here would only apply below its sound horizon \cite{Hu98}.

We allow an arbitrary but large range of the equation of state within $\Delta
w=4$ of the fiducial $w=-1$, so $-5\leq w\leq 3$. We limit the range of $w$ to
enable the MCMC sampling to converge to the joint posterior distribution of
the parameters more easily. This range is large enough to include extreme
departures from \lcdm\ and quintessence models, particularly considering the
conservative nature of our priors on PC amplitudes (see
Appendix~\ref{app:data}).  As with the quintessence model class, for the more
general class of smooth dark energy models we also examine the effects of
early dark energy and curvature on predictions for observables.

Figure~\ref{fig:dectree} shows a Venn diagram representation of the 
model classes that we study.
The \lcdm\ model class forms the innermost circle since it is 
a subset of all the other dark energy classes. Quintessence occupies a 
larger portion of the model space since the equation of state is allowed 
to vary within the range $-1\leq w\leq 1$, and models with even more 
general equations of state are contained within the outer 
``smooth dark energy'' circle.
The four quadrants of this diagram separate models that are either flat 
or have nonzero curvature and that either do or do not have a significant 
fraction of early dark energy. Within each of the three model classes, an 
open circle marks the simplest type of model (flat with no early dark energy), 
and arrows lead to other models within the class with additional 
degrees of freedom. 
Since the requirement that $w=-1$ at all times for \lcdm\ 
models implies that dark energy is always negligible at high redshift 
(see Fig.~\ref{fig:fid}), the \lcdm\ class has only two types of models instead of four.
See \S~\ref{sec:conc} for an index relating these model classes to the 
figures in the following sections.

The plots we present in the following sections show the range of fractional
deviations in observables relative to the fiducial flat \lcdm\ model that is
allowed by the assumed SN and CMB data, based on the distribution of MCMC
samples. We plot these predictions for various growth and expansion
observables (\S~\ref{sec:observ}) at redshifts $0\leq z\leq 4$. Any future
measurements that fall outside the predicted range of values would falsify a
model class.

Since it is impractical to show the whole 
posterior distribution at a number of redshifts for several different observables, 
we plot only the 68\% and 95\% limits of the distributions and a 
single example model selected from the MCMC samples.
The confidence limits are defined so that the probability (i.e.\ number of samples) 
is equal at the upper and lower limits, with 68\% (or 95\%) of the samples 
between those limits.  This definition corresponds to the 
``minimum credible interval'' (MCI) of Ref.~\cite{HHRW07}.
A useful property of these MCI limits is that the confidence region includes 
the mode of the samples even when the distribution is strongly skewed.

The redshift range in our plots of observable predictions extends beyond the 
coverage of the assumed SN sample so that we can make predictions for 
observables at higher redshift. Such high redshift observations are especially 
important for limiting the effects of curvature and early dark energy 
parameters that can change observables at redshifts beyond the reach of 
the SN data set, although the CMB distance prior can also play this role 
in simpler classes of models.

There are two ways in which a model class may still be falsified even if
future growth and expansion observations appear to be consistent with the SN
and CMB predictions for that class.  Observations that point to redshift
evolution of an observable that is inconsistent with the evolution in the
majority of samples would exclude the corresponding model class despite
appearing to be consistent with the predictions at any single redshift.
Similarly, measurements of multiple observables could falsify a model class if
they are inconsistent with the predicted correlations between those
observables.  Since these types of inconsistencies between predictions and
observations can be difficult to see in the types of plots shown here, we use
the example models in each figure to help point out some of the trends in
redshift and between observables.

In this study we do not address the feasibility of making the
measurements required to falsify various model classes using future data
sets. Instead, we focus here on determining what types of observables are the
most effective at distinguishing competing theories for acceleration and what
kind of precision in their measurement would be required.  We leave the task of
connecting our results with realistic expectations for upcoming dark energy
probes for future work.

%--------------------------------------
\subsection{Testing \lcdm}
\label{sec:tree1}

\begin{figure}[t]
\centerline{\psfig{file=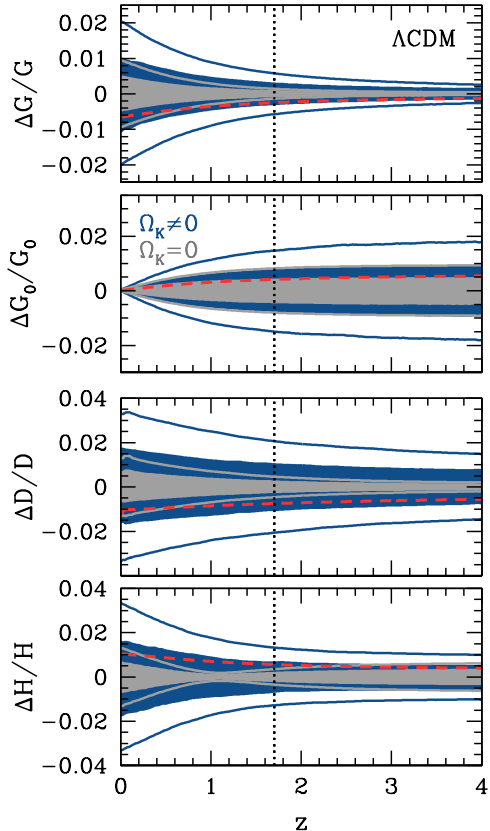, width=2.7in}}
\caption{Forecasted predictions from SNAP SN and Planck CMB data for growth
  and expansion  observables, showing the influence of curvature on 
predictions for \lcdm\ models.  The growth function $G$ is
  defined relative to its value at recombination, and $G_0$ is defined
  relative to the present value. Absolute distance $D$ differs from SN
  relative distances due to uncertainties in $H_0$.  
Shaded regions enclose 68\% CL 
regions and curves without shading are upper and lower 95\% CL limits, 
plotted as fractional differences from the
  fiducial flat \lcdm\ cosmology.  The model classes are flat \lcdm\ (\emph{light gray}) 
and \lcdm\ with nonzero spatial curvature (\emph{dark blue}).  An example model
with nonzero curvature is also shown (\emph{dashed red curve}).
Figures \ref{fig:zdist1_1}$-$\ref{fig:zdist2_3} and~\ref{fig:zdist2_4} all follow the same format.
  }
\vskip 0.25cm
\label{fig:zdist0}
\end{figure}

The flat \lcdm\ model has only two free parameters, $\om$ and $H_0$, whose
values are closely tied together by the CMB prior on $\omhh$.  With this
simple model, the fiducial SN and CMB data, or even the CMB data alone, make
strong predictions for the other observables (see Fig.~\ref{fig:zdist0}).  

The uncertainty in the growth function $G(z)$ at redshifts approaching
recombination is zero by definition, and only increases to $0.5\%$ by $z=0$
(quoting $68\%$ CL here and throughout this section). The expansion
observables at $\zmax$, $D(z=1.7)$ and $H(z=1.7)$, are predicted with $0.4\%$
and $0.2\%$ accuracy, respectively. At low $z$, $D$ and $H$ have equal
fractional uncertainties (since $\lim_{z\to 0} D(z)=z/H_0$), corresponding to
an accuracy of $0.7\%$ for $H_{0}$ \cite{Hu_standards}.  In comparison, 
current estimates of $H_0$ have uncertainties of
$3.8\%$ from WMAP alone and $1.9\%$ from 
combined WMAP, SN, and BAO measurements
\cite{Komatsu_2008}.  Note that there is an extremely tight and potentially
falsifiable prediction for $H$ at $z \sim 1$ of $0.09\%$ in flat \lcdm.  This
prediction is driven mainly by the tight CMB distance prior which effectively
reduces the remaining freedom in \lcdm\ to one parameter.  In a flat universe,
$D = \int dz/H(z)$, so allowed variations in $H^2 \propto \omhh(1+z)^3$ at
high $z$ must be compensated at low $z$ by an opposing variation in $H_0$ to
preserve $\dlss$.

% ****************************************
% ****************************************

Given these strong predictions, which are in large part already available from
current data, any future detection of deviations in growth, absolute distance,
or the expansion rate at the percent level at {\it
  any} redshift would provide evidence against flat \lcdm.
The predictions
are driven mainly by the CMB prior and hence the SN data themselves can be
viewed as a stringent test of flat \lcdm\ (see e.g.\ \cite{KLLS04}).
Conversely, testing the flat \lcdm\ predictions on other observables does not
depend strongly on having the SNAP SN data set in hand.

Future observations that rule out flat \lcdm\ would indicate a need for
additional complexity in the model.  Although the predictions in
Fig.~\ref{fig:zdist0} show what measurements would falsify flat \lcdm, they do
not indicate what kind of generalizations of the model would give such alternate
predictions.  For example, tight predictions of $\Delta D/D$ at 
$z>3$ may not be so interesting if there are no reasonable models
that can generate deviations from the flat \lcdm\ predictions there.  
Even in the
context of falsifying flat \lcdm\ it is important then to look at 
the predictions of an extended class of models.
Of the possible directions for generalizing the flat \lcdm\ model outlined 
in Fig.~\ref{fig:dectree}, we first examine the effects of including 
spatial curvature. We then consider the alternate option of allowing time 
variation of the equation of state in the following sections.

For \lcdm\ with curvature, growth and expansion predictions
 from the fiducial data are roughly a factor of 2 weaker than for flat
\lcdm\ but remain at the percent level (see Fig. \ref{fig:zdist0}).  The
impact of the SN data is much greater when we allow curvature to vary, since
the CMB constraints alone are no longer sufficient to fix the observables at
low redshifts.  The maximum uncertainty, at $z=0$, is $1.1\%$ for $G$
and $1.7\%$ for $H_{0}$ from $D$ and $H$.  The pivot point in $H$ at $z\sim 1$
also disappears, leading to the largest fractional change in the precision of
predictions.  In terms of measuring or constraining curvature under the \lcdm\
paradigm, the redshifts with the weakest predictions and the largest change
from flat \lcdm\ offer the most fruitful epochs for measurement.  Conversely,
the redshifts with the strongest predictions offer the best opportunity to
falsify the cosmological constant altogether.   As with flat \lcdm, with current constraints
these predictions weaken only by a factor of $\sim 2$ \cite{current}.

In summary,
under the assumption of \lcdm, with or without curvature,  
SN and CMB observations make firm predictions $\sim 1\%$ for growth 
and expansion observables at all redshifts. Future measurements that rule out a cosmological
constant as the source of cosmic acceleration would represent a
significant advance from a standpoint of fundamental physics.  To proceed 
beyond this point and determine the best observables and redshifts to target 
in order to distinguish among alternate models for acceleration, we
need to widen the model class again.

% ****************************************
\begin{figure}[t]
\centerline{\psfig{file=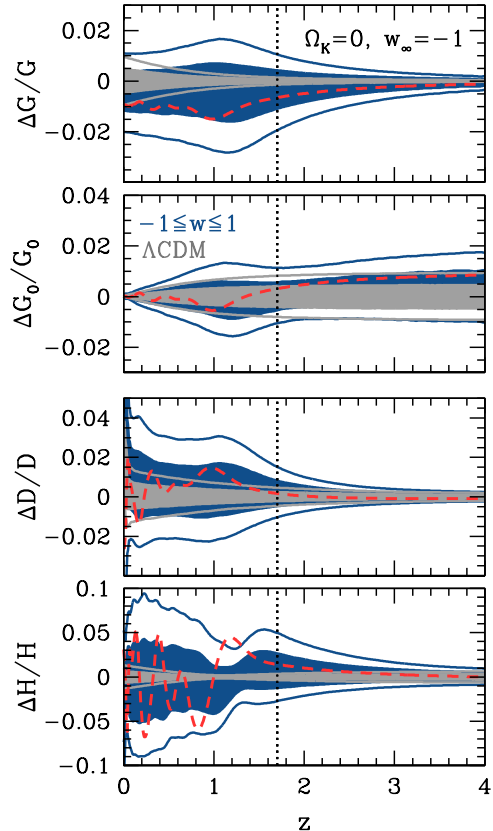, width=2.7in}}
\caption{{  Effects of generalization of flat \lcdm\ (\emph{light gray}) to quintessence (\emph{dark blue}; example model: \emph{dashed red}).}
Quintessence is defined to have 
$-1\leq w\leq 1$ with $w(z)$ parametrized by 15 PCs.
Here $\winf=-1$ and $\ok=0$ to eliminate early dark energy and curvature.
}
\vskip 0.25cm
\label{fig:zdist1_1}
\end{figure}
% ****************************************

%--------------------------------------
\subsection{Testing quintessence}
\label{sec:tree2}

If measurements of growth or expansion observables exclude \lcdm\ as a viable model,
the remaining dark energy model classes are ones where
the dark energy equation of state at late times ($z<\zmax$) is a 
free function of redshift (see Fig.~\ref{fig:dectree}). 
We add this freedom to the models by parametrizing
$w(z<\zmax)$ with PCs as described in \S~\ref{sec:pcs}.  
We use the first $15$ PCs of $w(z)$ at $z<\zmax$ in the MCMC 
likelihood analysis as this number suffices 
for completeness in the observables to percent level precision 
(see \S~\ref{sec:tests} and Appendix \ref{app:complete}).

As long as the scalar field potential remains positive, the equation of state
for quintessence is bounded in the interval $-1\leq w \leq 1$.  Negative
potentials that violate this bound in the past either would not produce the
required acceleration or would display easily falsifiable features.  Our
implementation of the canonical scalar field prior as described in
Appendix~\ref{app:data} is very conservative; all quintessence models are
allowed by the prior but not all models allowed by the prior can be
represented as a canonical scalar field.

We begin in Fig.~\ref{fig:zdist1_1} with the predictions for quintessence in a flat
universe with $w_\infty=-1$ to eliminate early dark energy.
Ruling out flatness in the context of \lcdm\ does not necessarily mean that
curvature is required in a more general model class, so we begin with the 
simplest quintessence models from Fig.~\ref{fig:dectree} and generalize 
to models with curvature and early dark energy later.

There are several notable features of quintessence predictions when compared
with those in the \lcdm\ class of models.  The limits on growth are no longer
monotonic with redshift and in particular show that growth suppression at the
$1-2\%$ level at $z \sim 1$, which would rule out \lcdm, is allowed in the more
general quintessence context.  Distances are typically predicted at a level
that is about twice that of flat \lcdm\ at $z<\zmax$ and 
comparable to \lcdm\ with
curvature.  The exception is at $z=0$ where the quintessence class
allows for sharp changes in the equation of state at $z < \zminsn$ as
described in Appendix~\ref{app:pcs}.  The prior on the Hubble constant
restricts the amplitude of such changes.  Conversely, with a prior on the
quintessence model class excluding these sharp transitions, precision Hubble
constant measurements would play the same role as 
absolute distance $D(z)$ or $H(z)$ measurements at $\zminsn < z \lesssim 0.1$
here and in all of the following cases.

Finally, the quintessence class allows a substantially wider range of Hubble
parameter predictions $H(z)$.  However, most of the allowed variation comes
from rapid oscillations in the dark energy density, demonstrated by the
example model in Fig.~\ref{fig:zdist1_1}.  As mentioned in
\S~\ref{sec:observ}, observations such as BAO that constrain $H$ are spread
over a wide redshift bin, averaging out much of this oscillatory behavior.
Constant shifts in the average $\Delta H/H \sim -\Delta D/D$ that 
would be allowed
by the SN measurements are still highly constrained in this model context
by the CMB distance measurement.  Without curvature or early dark energy, the
absolute distance between $\zmax$ and recombination is nearly fixed and thus
the Planck measurement also fixes shifts in the distance scale below $\zmax$
(see Appendix \ref{app:pcs}).

The flat quintessence class of models with no early dark energy on the whole remains
highly predictive and falsifiable.    We next examine what kind of observations might
falsify this class by requiring early dark energy or non-flat geometries.   

% ****************************************
\begin{figure}[t]
\centerline{\psfig{file=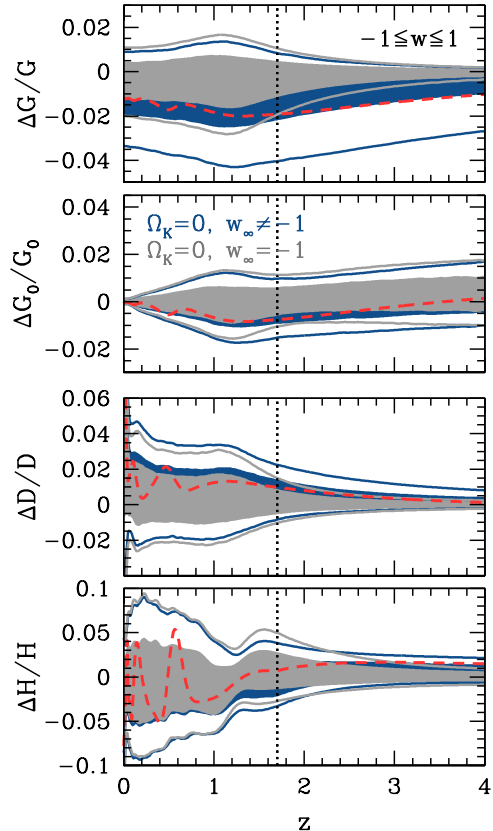, width=2.7in}}
\caption{{  Effects of early dark energy (\emph{dark blue}; example model: \emph{dashed red}) on quintessence models
(\emph{light gray}).}
Quintessence models from Fig.~\ref{fig:zdist1_1} are generalized to have $\winf$ 
vary, with $\ok=0$ to eliminate spatial curvature.   }
\vskip 0.25cm
\label{fig:zdist1_2}
\end{figure}
% ****************************************

% ****************************************
\begin{figure}[t]
\centerline{\psfig{file=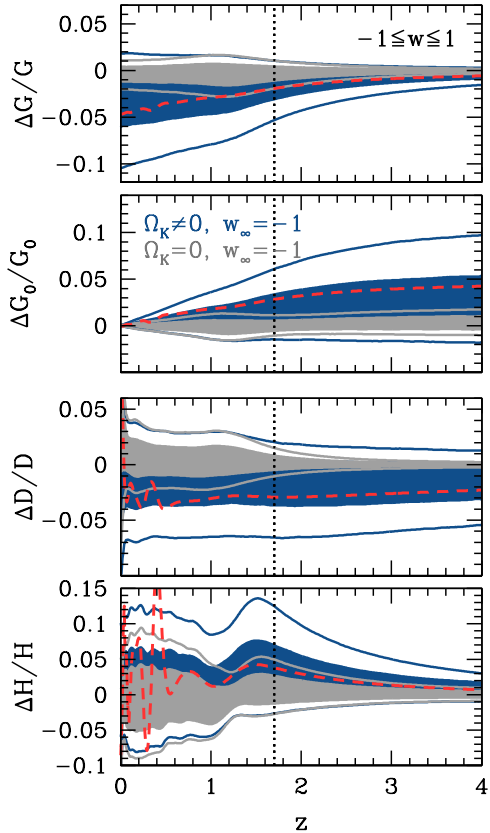, width=2.7in}}
\caption{{  Effects of curvature (\emph{dark blue}; example model: \emph{dashed red}) on quintessence models (\emph{light gray}).}
Quintessence models from Fig.~\ref{fig:zdist1_1} are generalized to have $\ok$ vary, with
$\winf=-1$ to eliminate early dark energy.   }
\vskip 0.25cm
\label{fig:zdist1_3}
\end{figure}
% ****************************************

% ****************************************
\begin{figure}[t]
\centerline{\psfig{file=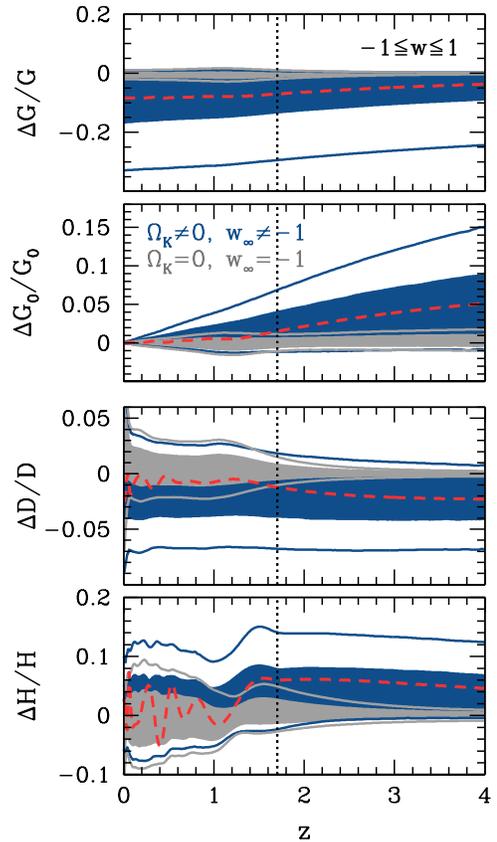, width=2.7in}}
\caption{{  Effects of curvature and early dark energy (\emph{dark blue}; 
example model: \emph{dashed red})  on quintessence models (\emph{light gray}).}
Quintessence models from Fig.~\ref{fig:zdist1_1} are generalized to have both $\ok$ and
$\winf$  vary.}
  \vskip 0.25cm
\label{fig:zdist1_4}
\end{figure}
% ****************************************

Fig.~\ref{fig:zdist1_2} shows the predictions for quintessence in a flat
universe with $\winf \ne -1$, i.e. with early dark energy.  Remarkably, the
predictions for absolute distance observables remain nearly unchanged.  In
particular, there is no substantial increase at $z \gtrsim \zmax$ nor are
compensating $H$ and $D$ shifts at $z<\zmax$ allowed.  
The lack of additional freedom in the observables is mainly due to
restricting the equation of state to $-1 \le w \le 1$.  Early dark energy
allows increased $H$ at $z> \zmax$ and hence decreased distance between
$\zmax$ and recombination.  To remain consistent with the 
CMB measurement of $\dlss$, these changes must be compensated by a reduction
in the average $\Delta H/H$ and an increase in absolute distances below
$\zmax$.  Given that the fiducial model is \lcdm\ with $w=-1$, a roughly
constant negative shift in $H$ requires dark energy to decrease with redshift,
i.e. a ``phantom'' equation of state with $w<-1$.  Since such values of $w$ 
are not allowed in the quintessence class, predictions
for $D$ and $H$ remain tight even with early dark energy.

On the other hand, although growth predictions are still at the $\sim 1-2\%$
level they show an interesting feature that is a signature of early dark
energy.  Growth suppression at $z \gtrsim \zmax$ is allowed at a larger level
and results in a nearly constant offset in growth at lower redshifts (see dashed curve in
Fig.~\ref{fig:zdist1_2}).  Positive $\Delta G/G$ is not allowed since 
the amount of early
dark energy can only increase from the fiducial model, which has almost no
early dark energy due to the assumption of $\winf = -1$.  Growth relative to
$z=0$, $\Delta G_0/G_0$, is largely unaffected by the extra freedom allowed by
early dark energy.  The smoking gun of early dark energy is therefore a
component to the growth function deviation that is nearly flat at $z \lesssim
\zmax$.

Fig.~\ref{fig:zdist1_3} shows the predictions for quintessence in a non-flat
universe with $\winf= -1$, i.e.\ with no early dark energy. Here the first
difference is the change in the {\it relative} growth $G_0$ between $\zmax$
and $z=0$ that is now allowed to be several percent.  While both early dark
energy and curvature can suppress growth relative to the early matter
dominated epoch, this shift in $G_0$ is a unique signature of nonzero
curvature.  A measurement that indicates $\Delta G_0/G_0 \sim 2-4\%$ 
at $z\sim \zmax$ would
falsify flat quintessence models, with or without early dark energy.  A
measurement beyond this level (or with the opposite sign) would falsify
non-flat cases as well.

The second main difference due to curvature is that high redshift negative
deviations in $D(z)$ at $z \gtrsim \zmax$ are now allowed at the $\sim 4\%$
level.  A small curvature affects the distance to recombination more than
it does distances at lower redshift.  In an open universe, $\dlss$ becomes larger and
therefore allows dark energy the freedom to compensate by introducing a
constant shift down in $\Delta D/D$ and up in $\Delta H/H$ at $z<\zmax$.  
The allowed
amplitude of the shift is limited by our BAO prior at $z=0.35$.  Note that
since the shift is constant in redshift, any single absolute distance
measurement at $z < \zmax$ also suffices to constrain this mode.  An example
of such a nearly constant shift is shown by the dashed curve in
Fig.~\ref{fig:zdist1_3}.

In a closed universe, the required compensation at $z<\zmax$ to preserve
$\dlss$ is in the opposite direction, and as with early dark energy 
the $w\geq -1$ quintessence prior limits
this possibility.  Thus one-sided deviations in $D$, average $H$, 
and growth relative to $z=0$ are signatures of
curvature in the quintessence model class.

Fig.~\ref{fig:zdist1_4} shows the predictions for quintessence with both
curvature and early dark energy.  Here a much greater range of early dark
energy densities is allowed since curvature in an open universe can 
compensate for the decreased distance to recombination due to early dark energy.
The main difference is a large increase in the allowed nearly constant offset
in the growth $G$ relative to recombination at $z \lesssim \zmax$ which can
now approach $20\%$.  The growth deviations are in fact limited by our early
dark energy prior from WMAP (see Appendix~\ref{app:data}); without this prior
the growth offset could have reached $\sim 40\%$.  Large suppression of the
growth relative to recombination indicates {\it both} early dark energy and
curvature in the quintessence context.

Note that the growth at $z<\zmax$ relative to $z=0$ remains nearly as well
predicted as in Fig.~\ref{fig:zdist1_3} with no early dark energy and hence is
equally falsifiable.  Likewise, predictions for $D$ do not weaken further
because constant deviations in distance are limited by the BAO prior on $D(z=0.35)$ in
both cases.  On the other hand, allowed models with significant early dark
energy do weaken predictions for $H$ at $z \gtrsim \zmax$.

Even in the most general quintessence class, there are still a few firm, 
percent level predictions.  
Neither the growth $G$ nor the distance $D$ can be appreciably {\it larger} than the \lcdm\ prediction, 
although both can be smaller.   
Lower average $H$ than in the fiducial flat \lcdm\ model is not allowed at $z<\zmax$ due 
to the $w\geq -1$ bound and at $z>\zmax$ due to the CMB prior on $\omhh$. 
Suppression of growth relative to high $z$ at a level of $\gtrsim 5\%$ 
at $z=\zmax$ must remain nearly constant at $z \lesssim \zmax$.  
Observations that violate these predictions would falsify the entire 
quintessence model class.

%--------------------------------------
\subsection{Testing smooth dark energy}
\label{sec:tree3}

Falsification of quintessence would challenge many theories of
dark energy and motivate consideration of more complicated models
than single canonical scalar fields.  Our generalization to the smooth dark
energy class encompasses equations of state with $-5\le w \le 3$ and requires
that the dark energy remain smooth compared with the matter on scales associated
with growth measurements.

\begin{figure}[t]
\centerline{\psfig{file=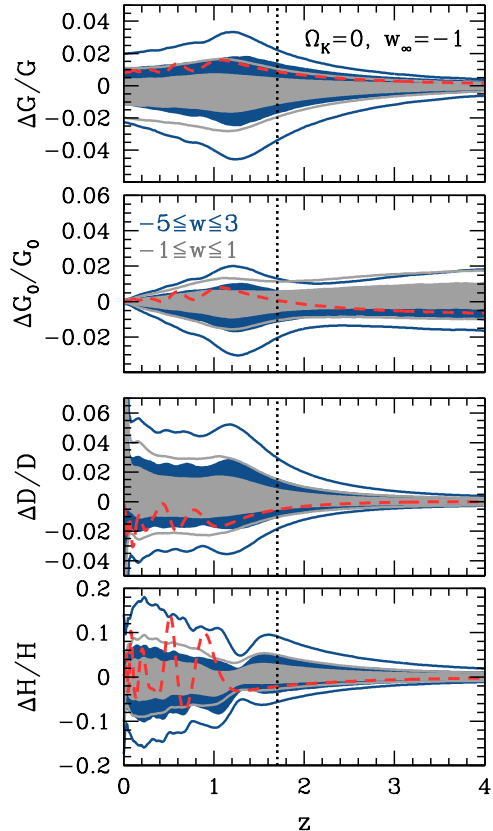, width=2.7in}}
\caption{{ Effects of generalizing $-1\leq w\leq 1$ quintessence models
    (\emph{light gray}) to smooth dark energy with $-5\leq w\leq 3$
    (\emph{dark blue}; example model: \emph{dashed red}).}  The smooth dark
  energy model class generalizes the quintessence models from
  Fig.~\ref{fig:zdist1_1} using the same 15 PCs for $w(z<\zmax)$ with
  $\winf=-1$ and $\ok=0$ to eliminate early dark energy and spatial
  curvature.}
\vskip 0.25cm
\label{fig:zdist2_1}
\end{figure}

\begin{figure}[t]
\centerline{\psfig{file=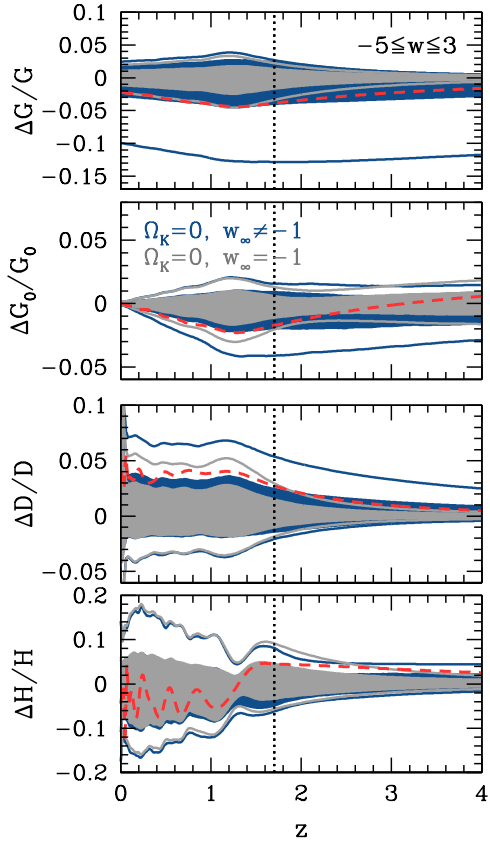, width=2.7in}}
\caption{{  Effects of early dark energy  (\emph{dark blue}; example model: \emph{dashed red}) 
 on smooth dark energy models (\emph{light gray}).}  Smooth dark energy models
 from Fig.~\ref{fig:zdist2_1} are generalized to have $\winf$ vary, with $\ok=0$ to eliminate 
 spatial curvature.
  }
\vskip 0.25cm
\label{fig:zdist2_2}
\end{figure}

\begin{figure}[t]
\centerline{\psfig{file=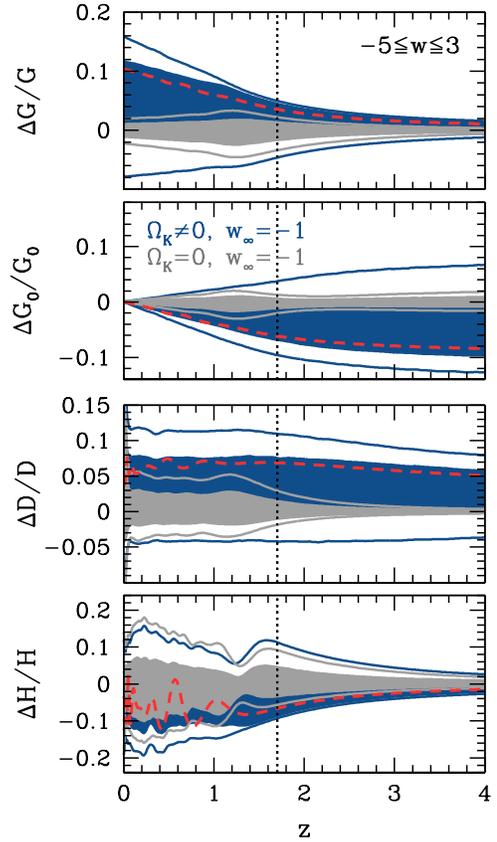, width=2.7in}}
\caption{{  Effects of curvature  (\emph{dark blue}; example model: \emph{dashed red})  on smooth
dark energy models (\emph{light gray}).}
 Smooth dark energy models
 from Fig.~\ref{fig:zdist2_1} are generalized to have $\ok$ vary, with $\winf=-1$ to eliminate 
early dark energy.}
\vskip 0.25cm
\label{fig:zdist2_3}
\end{figure}

\begin{figure}[t]
\centerline{\psfig{file=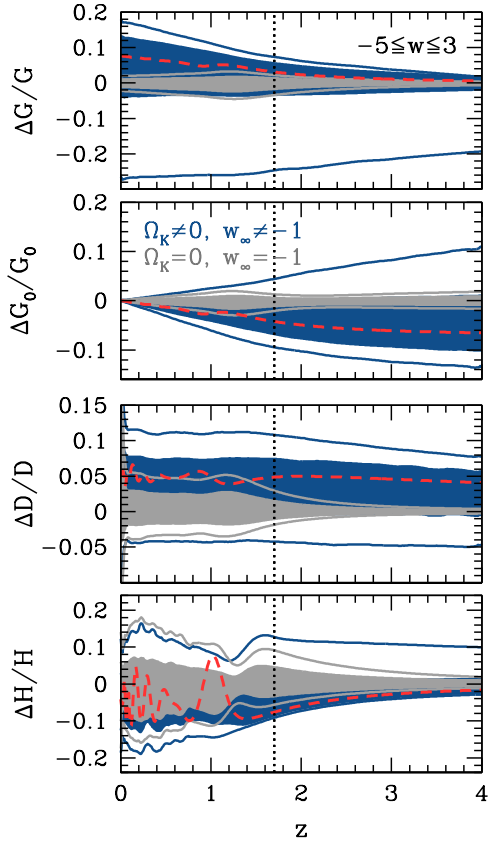, width=2.7in}}
\caption{{ Effects of curvature and early dark energy (\emph{dark blue};
    example model: \emph{dashed red}) on smooth dark energy models
    (\emph{light gray}).}  Smooth dark energy models from
  Fig.~\ref{fig:zdist2_1} are generalized to have both $\winf$ and $\ok$
  vary.  }
\vskip 0.25cm
\label{fig:zdist2_4}
\end{figure}

Figures~\ref{fig:zdist2_1}$-$\ref{fig:zdist2_3} and~\ref{fig:zdist2_4} show the growth and expansion
predictions from SN and CMB data 
relative to the fiducial model for the class of general smooth
dark energy models.  As with quintessence, we present predictions for models
both with and without curvature and/or early dark energy.  Remarkably,
Figure~\ref{fig:zdist2_1} shows that for flat models without early dark energy
the effect of dropping the quintessence bounds on $w(z)$ weakens 
predictions by less than a factor of two.  Thus the more
general class of smooth dark energy without curvature or early dark energy is
nearly as falsifiable as flat quintessence.

% ****************************************
% ****************************************

% ****************************************
% ****************************************

Including early dark energy in smooth dark energy models relaxes the
predictions in ways qualitatively similar to the quintessence case, but with
allowed deviations that are somewhat larger at 68\% CL and noticeably larger
at 95\% CL.  The additional freedom in growth and distances comes mainly from
dropping the lower bound on $w(z)$ and allowing phantom dark energy models
with $w<-1$. This change enables a reduction of the dark energy density 
at $z<\zmax$ from the fiducial $w=-1$ model, which can compensate for and thereby
allow a higher fraction of early dark energy while maintaining consistency with
the CMB distance prior.  Since dark energy is the dominant component at late
times, decreasing its density typically results in lower total density at low
$z$, which increases absolute distances there. The extra early dark energy
suppresses the growth at $z>\zmax$, and the reduced dark energy density at
low $z$ raises growth back up toward the fiducial model slightly by $z=0$.  
Comparing Figs.~\ref{fig:zdist1_4} and~\ref{fig:zdist2_2}, we find that these models
with significantly lower $H(z<\zmax)$ are the only flat smooth dark energy
models that would not already be excluded if observations had falsified all
quintessence models.

% ****************************************
% ****************************************

In contrast, allowing spatial curvature to vary in the smooth dark energy
model class produces models with qualitatively new types of behavior not seen
in the quintessence models.  These new observable deviations from flat \lcdm\ 
provide smoking-gun signatures of curvature and an
equation of state beyond quintessence.

Although some of the model classes examined so far allow growth relative to early times to 
be suppressed by $\sim 20\%$ or more compared with the fiducial flat \lcdm\
model, none of the previous cases allow growth to be enhanced by more than $\sim 2\%$.  However,
models with weak bounds on $w(z)$ and nonzero spatial curvature can have
$G(z)$ be as much as $10-15\%$ higher than the fiducial model at $z=0$, as
shown in Figure~\ref{fig:zdist2_3}.  Likewise, the growth $G_0$ relative to
$z=0$ can be lower than in the fiducial model by $5-10\%$ or more at $z\gtrsim
\zmax$, whereas it was previously limited to deviations of at most a few
percent in this direction.  Both effects correspond to an enhancement of
growth at $z\lesssim \zmax$ that occurs in closed models ($\ok < 0$).

The reason why such models become viable when we abandon the $-1\leq w\leq 1$
prior is similar to the explanation for the differences in the early dark
energy predictions of quintessence and smooth dark energy models.  Closed
universes have smaller distances to recombination, so without some other means
to increase the distance, the models with $\ok<0$ are inconsistent with the CMB
distance prior for the fiducial cosmology. Removing the quintessence bounds on
$w(z)$ allows for lower dark energy density (with $w<-1$) at low $z$, which
increases the total distance to last scattering and allows closed models to
match the CMB constraints. The lower dark energy density at $z<\zmax$ and
resulting enhancement of distances are also reflected in the predictions for
$H$ and $D$ in Fig.~\ref{fig:zdist2_3}. Having $\ok<0$ and $\ode(z<\zmax)$
lower than the fiducial value means that $\om(z<\zmax)$ is higher than in the
fiducial model, which also contributes to the additional growth of structure
at low redshift.

Although the average $H(z)$ at low redshift can be reduced considerably 
relative to flat \lcdm, predictions for $H(z)$ at high redshift
have a sharp lower limit. The CMB constraint 
on $\omhh$ places a strong lower bound on $H(z)$ corresponding to the expansion 
rate in an Einstein-de Sitter universe. Closed models can exceed this bound slightly 
since the curvature slows the expansion, but for allowed values of $\ok$ 
this is a small effect.  This lower limit is present in all previous model classes as well but
its impact is less visible.

% ****************************************
\begin{figure}[t]
\centerline{\psfig{file=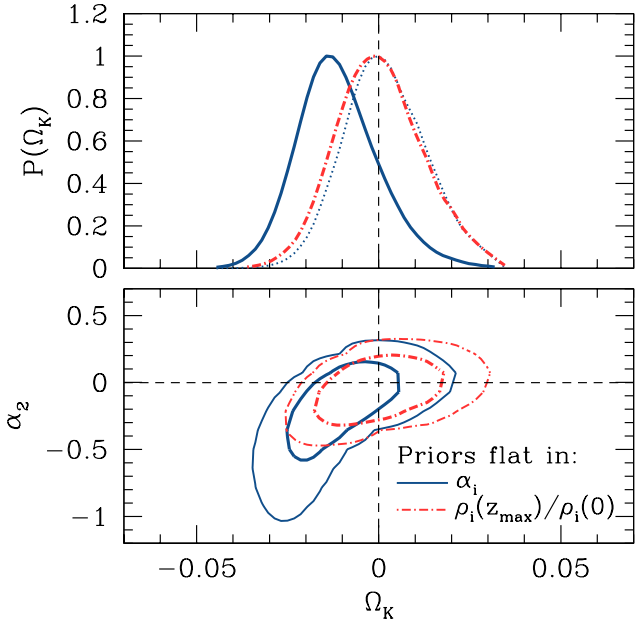, width=3.5in}}
\caption{Effects of priors on the degeneracy between curvature and $w(z)$ PCs 
for the smooth dark energy models with curvature 
(but not early dark energy) in Fig.~\ref{fig:zdist2_3}.
\emph{Top}: 1D marginalized posterior probability $P(\ok)$ 
for the default priors that are flat in PC 
amplitudes $\alpha_i$ (\emph{solid, dark blue}), and for alternate priors that are
flat in the density of each PC at $\zmax$ relative to $z=0$, 
$\rho_i(\zmax)/\rho_i(0)$ as defined in Eq.~(\ref{eq:flatrhoprior}) 
(\emph{dot-dashed, red}). The dotted curve is the 
mean likelihood distribution for flat $\alpha_i$ priors.
\emph{Bottom}: Probability contours of $\ok$ vs.\ $\alpha_2$ 
at 68\% and 95\% CL for the same priors on PCs as in the top panel.
Dashed lines mark the fiducial values, $\ok=0$ and $\alpha_2=0$.
}
\vskip 0.25cm
\label{fig:omkflatrho}
\end{figure}
% ****************************************

% ****************************************
\begin{figure}[t]
\centerline{\psfig{file=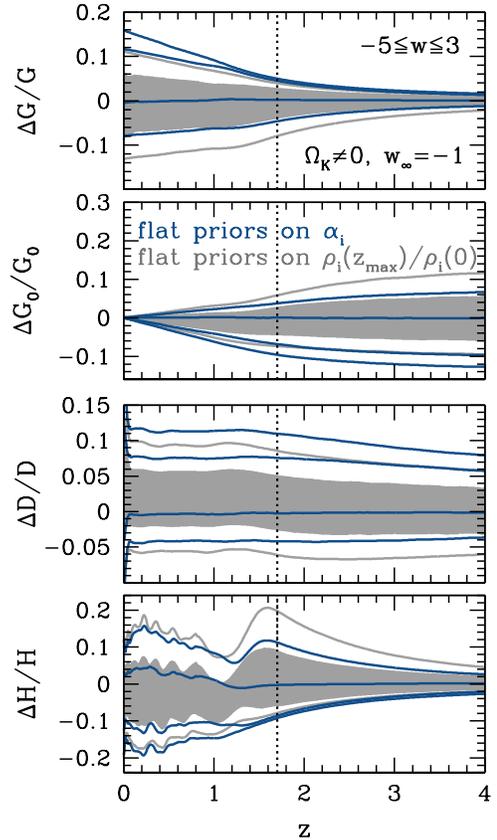, width=2.7in}}
\caption{Effects of priors on smooth dark energy models with nonzero curvature.
 \emph{Dark blue}: flat top-hat priors on $\alpha_i$ as in Fig.~\ref{fig:zdist2_3}.
 \emph{Light gray}: flat priors on the density of individual PCs, $\rho_i(\zmax)/\rho_i(0)$.
}
\vskip 0.25cm
\label{fig:flatrho1}
\end{figure}
% ****************************************

Like the non-flat quintessence predictions of Fig.~\ref{fig:zdist1_3}, 
the predictions for smooth dark energy models with curvature 
are asymmetric about the fiducial model. In fact, 
as Fig.~\ref{fig:zdist2_3} shows, the confidence regions 
are so skewed toward the closed models that
the fiducial model lies on the edge of the 68\% regions.  This is in spite of
the fact that the fiducial model has the maximum likelihood by definition, and
that the definition of confidence limits we use is chosen to include the peak
probability.  Moreover, the allowed regions of the observables in
Fig.~\ref{fig:zdist2_3} are skewed in the opposite direction of the
predictions for the corresponding quintessence models in
Fig.~\ref{fig:zdist1_3} and therefore appear inconsistent with those
predictions, despite the fact that the quintessence class is a subset of the
more general smooth dark energy class.  The reason for these discrepancies is
that predictions in the non-flat cases with large variations in $w(z)$ both
above and below $w=-1$ are so weak that the shapes of priors on the PC
amplitudes become important in determining the extent of the confidence
regions.

The influence of priors on the predictions is illustrated in
Fig.~\ref{fig:omkflatrho}, where we show distributions of $\ok$ for two
different choices of priors. With our usual top-hat prior on $\{\alpha_i\}$, the
posterior probability for $\ok$ is strongly skewed toward closed models. The
distribution of the mean likelihood of MCMC samples, on the other hand, is
peaked at the fiducial value of $\ok=0$ as expected (dotted curve in
Fig.~\ref{fig:omkflatrho}).

The discrepancy between posterior probability and
mean likelihood can be traced to a volume effect in the parameter space
(e.g. see~\cite{Lewis:2002ah}).  Models with more negative $\ok$ have a wider
range of values of $\alpha_i$ that fit the data well; this is demonstrated by
the banana-shaped contours for $\ok$ and $\alpha_2$ in the lower panel of
Fig.~\ref{fig:omkflatrho}, and other PCs show similar widening of the
parameter volume at more negative $\ok$.  When other parameters are
marginalized to obtain the 1D posterior distribution for $\ok$, or for one of the
growth and expansion observables at some redshift, the result is a skewed
distribution.

The basic reason for this volume effect is that the dark energy 
density depends exponentially on $w(z)$, which is a linear 
combination of the PCs, so changes in $\{\alpha_i\}$ at small $\rhode$ 
have less effect on observables than changes at large $\rhode$. 
To test how much the observable predictions are affected by the priors, 
we use alternate priors that are flat in the contribution of each principal component 
to the dark energy density at $\zmax$ relative to $z=0$
(see Appendix~\ref{app:data}),
\begin{equation}
\frac{\rho_i(\zmax)}{\rho_i(0)} \equiv \exp\left[3\alpha_i
\int_0^{\zmax} dz \frac{e_i(z)}{1+z}\right],
\label{eq:flatrhoprior}
\end{equation}
so that the total dark energy density at $\zmax$ is $\rhode(\zmax) = \rhode(0)
\prod_i [\rho_i(\zmax)/\rho_i(0)]$.  Figure~\ref{fig:flatrho1} shows that with
this new prior, the predictions become more symmetric around the fiducial
model and also allow models that are acceptable under the quintessence
subclass.

These $\sim 1 \sigma$ shifts indicate that the predictions from cosmological
data alone are so weak that the exact confidence region depends on arbitrary
theoretical priors on the measure in dark energy model space.  Therefore, any
measurement of these observables at a level of precision comparable to the
predictions is interesting in the context of smooth dark energy with curvature
regardless of the sign of the deviation from flat \lcdm.  At the same time, conclusive
falsification of general smooth dark energy models with curvature would
require much larger deviations where there are no models with good likelihood
values.

Note that the dependence on PC priors is {\it only} significant for classes of models 
that allow  $w<-1$ and have nonzero curvature. For all 
of the previous cases -- \lcdm, quintessence, and $\ok=0$ smooth dark energy -- the 
confidence limits of observables shift by only $\lesssim 1\%$ when we switch 
from one set of PC priors to the other (except for $H$, for which the limits change 
by up to a few percent at some redshifts near $\zmax$).
The volume effect is not a consequence of our particular parametrization of $w(z)$; 
for example, there is a similar shift toward $\ok<0$ for non-flat dark energy models 
parametrized as $w(a)=w_0 + w_a(1-a)$ when the priors are flat in $w_0$ 
and $w_a$ \cite{current}.

% ****************************************
% ****************************************

% ****************************************
\begin{figure}[t]
\centerline{\psfig{file=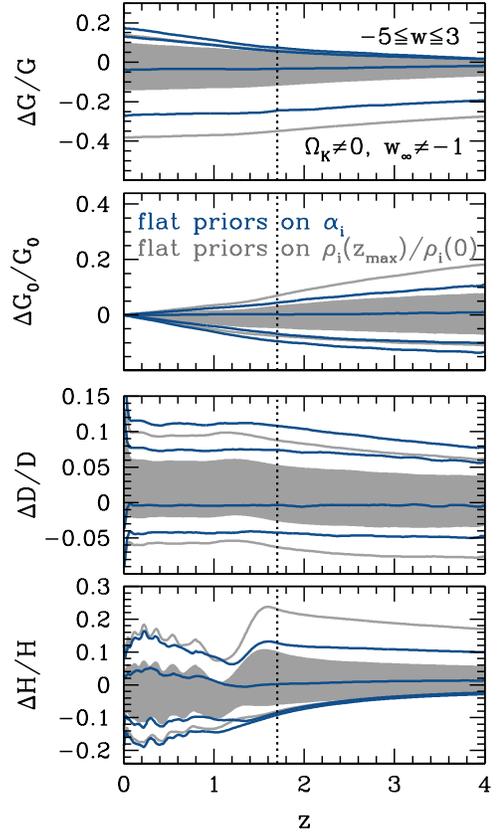, width=2.7in}}
\caption{{  Effects of priors on smooth dark energy models with curvature and early dark energy.}
Same as Fig.~\ref{fig:flatrho1}, but comparing priors for the dark blue model in 
Fig.~\ref{fig:zdist2_4}.
}
\vskip 0.25cm
\label{fig:flatrho2}
\end{figure}
% ****************************************

Using the priors that are flat in the PC amplitudes, the addition of early dark energy 
to smooth dark energy models with curvature (Fig.~\ref{fig:zdist2_4}) appears 
to make little difference to the qualitative predictions, except in some of 
the 95\% limits.  However, much of the effect of early dark energy on these models 
is masked by the flat-$\{\alpha_i\}$ PC prior. As Fig.~\ref{fig:flatrho2} shows, 
changing the priors to be flat in $\rho_i(\zmax)/\rho_i(0)$
affects the observable predictions in ways that are similar to the previous 
case without early dark energy, e.g. making many of the distributions of allowed models 
more symmetric around the fiducial model. 
With this alternate prior we see that the additional freedom in early dark energy 
combined with nonzero curvature 
enables models with significant growth suppression relative to high $z$ and increased 
$H(z \gtrsim \zmax)$ to fit the assumed data sets, as it does in the 
corresponding quintessence predictions of Fig.~\ref{fig:zdist1_4}.

The sensitivity to the priors and the weakness of the predictions in general
means that despite the great potential of future observations for
{\it measuring} spatial curvature and early dark energy in models with general
equation of state variation at low redshift, statements about 
{\it falsification} of
the entire smooth dark energy model class must be made with care.

%--------------------------------------
\subsection{Beyond Smooth Dark Energy}
\label{sec:tree4}

One lesson from the analysis in the previous section is that the combination 
of general variation in the dark energy equation of state with early dark energy
and nonzero spatial curvature allow a wide variety of cosmological models 
to fit future SN and CMB data. Falsification of 
the most general smooth dark energy model class therefore 
appears to be quite difficult,
especially given the dependence of the growth and expansion observable predictions 
on priors that must be set arbitrarily in the absence of 
an underlying theory for dark energy.
However, even these very general models make some firm predictions about the 
relations between observables that could potentially be falsified by 
future measurements.

One example of a robust prediction for the observables is the following: given the
flat \lcdm\ model that matches the SN and CMB data and assuming that the dark energy
always contributes positively to energy density, $H(z)$ at $z\gtrsim 2$ can
be no more than $\sim 5\%$ lower than in \lcdm.  This limit was noted in 
\S~\ref{sec:tree3}) in the context of non-flat smooth dark energy models 
with $\winf=-1$, and it still holds when we include early dark energy.

% ****************************************
\begin{figure}[t]
\centerline{\psfig{file=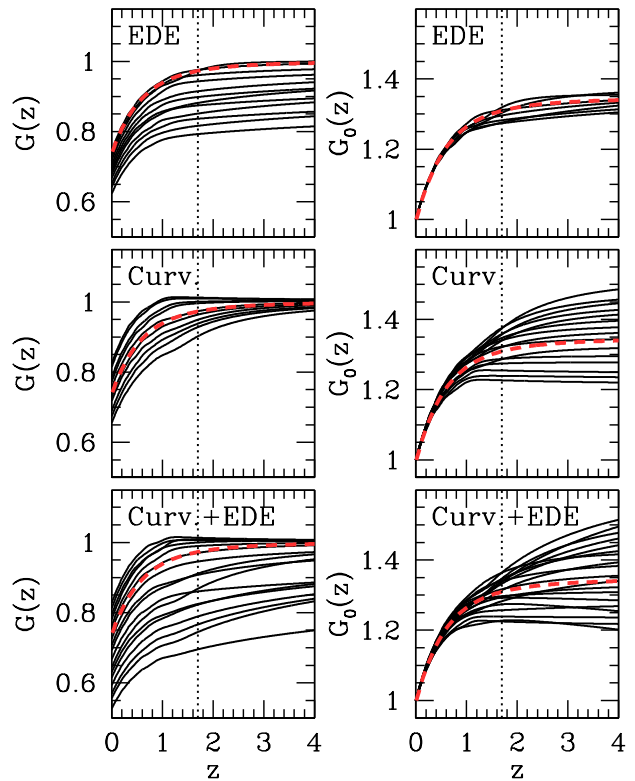, width=3.5in}}
\caption{Growth functions of MCMC samples 
in the smooth dark energy model class ($-5\leq w \leq 3$) that include 
either early dark energy (EDE) at $z>\zmax$ ($\winf \ne -1$; \emph{top}), 
curvature ($\ok\ne 0$; \emph{middle}), or both (\emph{bottom}). 
We plot growth relative to 
early times in the left column of panels, and growth relative to the present 
on the right.
Dashed red curves show growth in the fiducial flat \lcdm\ model.
Samples are selected randomly from those with likelihoods 
satisfying $\Delta \chi^2 \leq 4$, but for visual clarity we plot 
samples that are approximately evenly spaced in $G(z=0)$ (\emph{left}) 
or $G_0(z=4)$ (\emph{right}).
The dotted vertical line in each panel marks $z=\zmax$.
}
\vskip 0.25cm
\label{fig:gzmax}
\end{figure}
% ****************************************

More interestingly, the predicted redshift evolution of expansion and growth 
observables still exhibits certain regularities.
Notice in Fig.~\ref{fig:zdist2_2} that before we introduce curvature, 
the growth history, absolute distances, and expansion rate are predicted 
at the few percent level by SN and CMB data. Freedom in the spatial curvature 
greatly reduces the precision of these predictions. However, the curvature is 
set by a single parameter $\ok$ with well defined effects on each of the 
observables.  By taking advantage of our knowledge of the impact of 
curvature, we can effectively regain much of the predictive power that 
exists for flat models.

As an example, consider the growth function. In the most general model class, 
the majority of the freedom in growth comes from curvature and early 
dark energy. We can distinguish between the two by noting that the deviations 
in growth from the fiducial flat \lcdm\ model have different redshift 
dependence, as illustrated by the sample growth histories allowed by either 
curvature or early dark energy plotted in Fig.~\ref{fig:gzmax}.  
The main effect of early dark energy is to suppress growth by a constant factor 
at early times, so growth functions in a flat universe with varying 
amounts of early dark energy have similar shapes but different amplitudes at low $z$.
On the other hand, curvature tends to have a more gradual effect on growth 
continuing to $z=0$, and also allows more enhanced growth relative to the fiducial 
model than early dark energy.

% ****************************************
\begin{figure}[t]
\centerline{\psfig{file=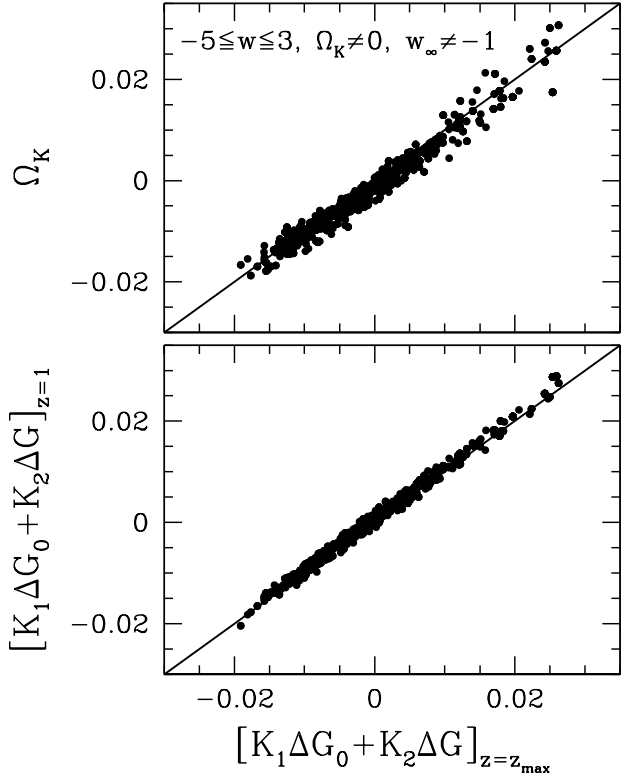, width=3.5in}}
\caption{Comparison of $\ok$ and linear combinations of 
the growth function relative to flat \lcdm, 
$\Delta G_0(z)$ and $\Delta G(z)$, at $z=\zmax$ and $z=1$ 
for randomly selected models (with $\Delta\chi^2\leq 4$) in the 
smooth dark energy class with early dark energy and curvature.
The coefficients used here are $K_1(z=1)=0.24$, $K_2(z=1)=0.10$, 
$K_1(z=\zmax)=0.17$, and $K_2(z=\zmax)=0.09$.
Note that these specific values may not produce accurate estimates 
of $\ok$ for fiducial cosmologies other than the one assumed here.
}
\vskip 0.25cm
\label{fig:gzmax2}
\end{figure}
% ****************************************

An observed growth history that cannot be described by some combination 
of the effects of curvature and early dark energy
would present a major challenge to the 
dark energy paradigm.  For example, growth at $z<\zmax$ relative to 
today ($G_0$) that is 
$\gtrsim 5\%$ higher than expected in flat \lcdm\ would be difficult 
to explain with dark energy for allowed values of the spatial curvature.  

Moreover, deviations in $G_0(z)$ are nearly a one parameter family
that is ordered by curvature $\ok$.   The top panel of 
Fig.~\ref{fig:gzmax2} shows the correlation
between $\ok$ and a linear combination of $G_0(\zmax)$ and $G(\zmax)$ 
in smooth dark energy models.  The curvature $\ok$ mainly depends on 
$G_0(\zmax)$, but the degeneracy in $\dlss$ between 
$\Delta H/H$ at $z<\zmax$ and early dark energy introduces a small 
correction since such changes in the expansion rate affect the growth rate 
at low $z$ and therefore change $G_0$. The $G(\zmax)$ term can correct for 
this degeneracy since the amplitude of $G$ is sensitive to the amount 
of early dark energy.

By comparing the combination of $G_0$ and $G$ correlated with curvature 
at multiple redshifts, one can test the general class of smooth dark 
energy models with early dark energy and curvature. 
For example, the lower panel of Fig.~\ref{fig:gzmax2} shows 
these combinations of growth observables at $\zmax=1.7$ and at $z=1$; 
observations that give different values for the linear combinations of $G_0$ 
and $G$ at these two redshifts would falsify this most general model class.

\begin{figure}[t]
\centerline{\psfig{file=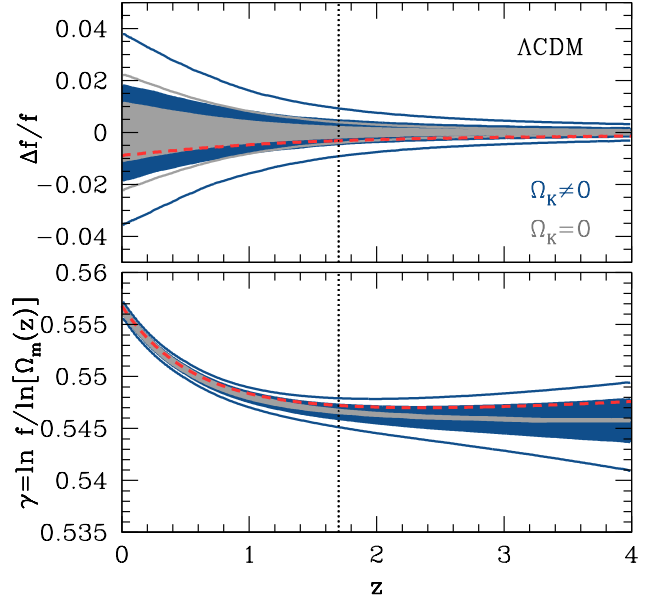, width=3.5in}}
\caption{Effects of curvature on the \lcdm\ growth rate.
Predictions from future SN and CMB data for the growth rate 
$f = 1 + d\ln G/d\ln a$, plotted relative to the fiducial model, 
and for the growth index $\gamma = \ln f / \ln[\om(z)]$. The model 
classes here are \lcdm\ either assuming flat geometry (\emph{light gray}) 
or allowing nonzero curvature (\emph{dark blue}; example model: 
\emph{dashed red}). Shaded regions enclose 68\% of the allowed models 
and curves without shading are upper and lower 95\% confidence limits 
(not visible for flat \lcdm\ predictions of $\gamma$ due to the 
tightness of those constraints).
The three example models plotted here and in Figs.~\ref{fig:gamma2} 
and~\ref{fig:gamma3} are the same as the ones 
in Figs.~\ref{fig:zdist0}, \ref{fig:zdist1_4}, and~\ref{fig:zdist2_4}, 
respectively.
}
\vskip 0.25cm
\label{fig:gamma1}
\end{figure}

\begin{figure}[t]
\centerline{\psfig{file=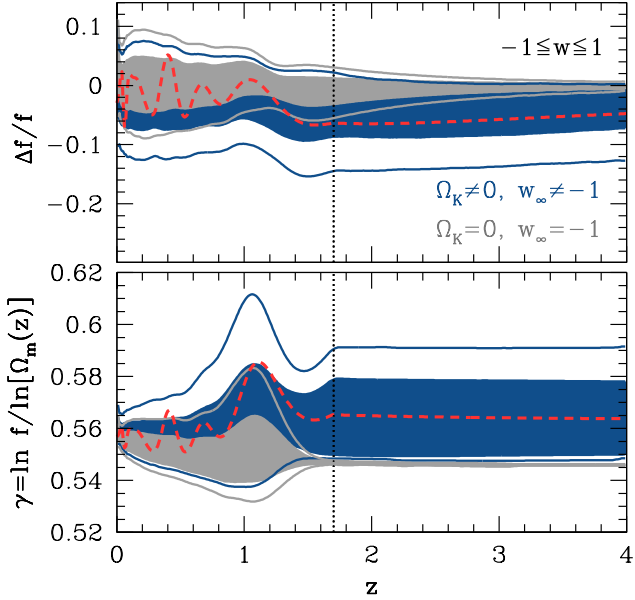, width=3.5in}}
\caption{Effects of curvature and early dark energy on the quintessence growth rate.
Same as Fig.~\ref{fig:gamma1} but for flat quintessence models 
with $\winf=-1$, i.e. no early dark energy (\emph{light gray}), and 
quintessence with both nonzero curvature and early dark energy 
(\emph{dark blue}; example model: \emph{dashed red}).
}
\vskip 0.25cm
\label{fig:gamma2}
\end{figure}

\begin{figure}[t]
\centerline{\psfig{file=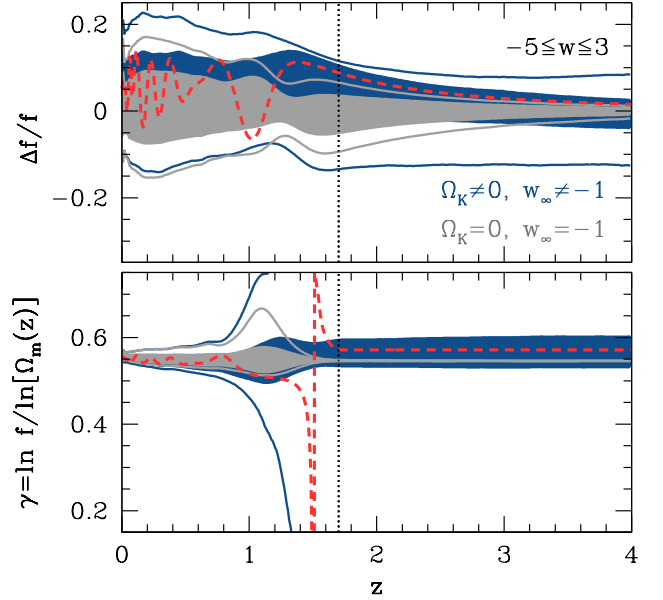, width=3.5in}}
\caption{Effects of curvature and early dark energy on the 
smooth dark energy growth rate.
Same as Fig.~\ref{fig:gamma2} but for smooth dark energy models 
($-5\leq w\leq 3$).
}
\vskip 0.25cm
\label{fig:gamma3}
\end{figure}

This regularity in the growth relative to today can be viewed as a generalization of
tests involving the linear growth rate $f(z)$.  
In particular, the relationship $f(z) = \om^{\gamma}(z)$ with
$\gamma\approx 0.55$ has been proposed as a potential
test of all smooth dark energy models \cite{Linder_gamma,Hut_Lind_MMG,LinCah07,PolGan08,Acqetal08}.
In Figs.~\ref{fig:gamma1}$-$\ref{fig:gamma3}, we plot the predictions 
from the forecasted SNAP supernovae and Planck CMB data for the 
growth rate and growth index for selected model classes from 
the previous sections, allowing a redshift dependent growth index
\begin{equation}
\gamma(z) = \frac{\ln[f(z)]}{\ln[\om(z)]}.
\label{eq:gammadef}
\end{equation}
Note that the growth index is somewhat different from our other observables 
since measurement of $\gamma$ requires not only the growth rate but 
also some method for determining the fractional matter density $\om(z)$ 
at the same redshift.  A measurement of $H(z)$ combined with the CMB 
constraint on $\omhh$ could provide an estimate of the latter quantity, since 
$\om(z)\propto \omhh (1+z)^3 H^{-2}(z)$.

In the context of \lcdm,
Figure~\ref{fig:gamma1} shows that both $f$ and $\gamma$ are precisely 
predicted by future SN and CMB data  as 
expected given the tight constraints on other observables within 
this model class. The growth index is nearly constant, with a small slope 
at low redshifts when the cosmological constant dominates, and 
deviations of $\sim 0.5\%$ by $z=4$ for models with nonzero curvature.

% ****************************************
% ****************************************

In the context of quintessence, both predictions weaken substantially 
as shown in Fig.~\ref{fig:gamma2}.
The growth {\it rate} is not as well predicted as the
difference in growth between $z=0$ and $\zmax$ used in Fig.~\ref{fig:gzmax2}.
Like $H$, it effectively has only one integral
over the time-varying equation of state instead of two as for 
distances and the integrated growth history.  It is therefore
equally sensitive to features in $w(z)$.

% ****************************************
% ****************************************

% ****************************************
% ****************************************

In fact, the redshift dependence of the 
predictions for $f(z)$ in Fig.~\ref{fig:gamma2} closely 
mimic those for $H(z)$ in Fig.~\ref{fig:zdist1_4}, but with opposite sign.
Since $\om(z)$ is 
tied to $H(z)$ via the CMB prior on $\omhh$, the similarity between 
deviations in $f$ and in $H$ suggests a strong connection between $f$ and $\om(z)$, 
consistent with a constant value of $\gamma$.
However, the predicted values of $\gamma$ shown in the lower panel 
of Fig.~\ref{fig:gamma2} cover a much wider range for quintessence 
than for \lcdm, with significant variation of $\gamma$ with redshift 
in some models. For flat models without early dark energy, 
the extra freedom in $\gamma$ only appears at $z<\zmax$, but 
including curvature and early dark energy allows $5\%$ deviations 
in $\gamma$ above the fiducial value at $z>\zmax$, only slightly less than 
the uncertainty in $f$ or $H$ at these redshifts. 
These nearly constant deviations in $\gamma$ at high $z$ are well 
approximated by the expected dependence on early dark energy given 
by Ref.~\cite{LinCah07}, modified for nonzero curvature,
\begin{equation}
\gamma(z>\zmax) \approx \frac{3(1-\winf)}{5-6\winf}f_{\rm DE} 
+ \frac{4}{7}f_{\rm K},
\label{eq:gammaede}
\end{equation}
where $f_{i} = \Omega_{i}(z)/[1-\om(z)]$. For $\ok=0$ and $\winf=-1$ 
we recover the usual growth index from this formula, $\gamma = 6/11 \approx 0.55$.

One of the more interesting features of the predictions for $\gamma$ in the
quintessence class is the widening in the uncertainty at $z\sim 1$. 
As Eq.~(\ref{eq:gammaede}) shows, deviation from $w=-1$  
changes the value of $\gamma$ at high redshift. Likewise, low-redshift
variation in $w$ can perturb the growth index from its usual value of 
$\gamma\approx 0.55$. At $z\lesssim 0.5$, SN distance constraints ensure 
that while $w$ may oscillate rapidly with redshift, it never deviates 
far from $w=-1$ on average. However, as $z$ approaches $\zmax$ the 
constraints from SNe weaken, allowing $w$ to vary significantly from $-1$ 
over longer periods of time; this variation enables $\gamma$ to 
deviate from the \lcdm\ prediction.  
Furthermore, at $z\gtrsim 1$ where $\om(z)$ is typically near unity,
$\ln[\om(z)]$ is close to zero and therefore the value of $\gamma$ derived from 
Eq.~(\ref{eq:gammadef}) is more sensitive to small changes in the 
relation between $f(z)$ and $\om(z)$.
For example, the model plotted in Figs.~\ref{fig:zdist1_4} and~\ref{fig:gamma2} 
has significant changes in $H$ and $f$ between $z=1$ and $z=1.5$ relative 
to the fiducial flat \lcdm\ cosmology, 
leading to a bump in $\gamma$ for this model at $z\sim 1$.

Generalizing to smooth dark energy models with $-5\leq w\leq 3$ 
reveals another potential problem with using $\gamma$ to test 
smooth dark energy. Recall from the previous section that a closed universe is 
allowed for models in this class, and is 
favored if priors are flat in the PC amplitudes.
In closed models where $\ok$ is sufficiently negative, $\om(z)$ 
can cross unity with $\om(z)>1$ at high $z$ and $\om(z)<1$ at low $z$, 
and the same is true of $f(z)$. Since these two functions do not 
cross at exactly the same redshift due to the slight lag between 
density and growth, the growth index of Eq.~(\ref{eq:gammadef}) has a 
singularity when $\om(z)=1$. The dashed curve in Fig.~\ref{fig:gamma3} 
is one example of such a model. As a result, the predictions for 
$\gamma$, particularly in the tails of the distribution, blow up 
at $z\gtrsim 1$ where these singularities occur.
This effect is an artifact of the $\gamma$ parametrization since 
$\gamma$ can take any value when $f\approx 1$ and $\om(z)\approx 1$, 
but it makes it difficult to interpret limits on $\gamma$ beyond $z\sim 1$
for the most general class of smooth dark energy models.

Although constant $\gamma$ remains a good approximation for many dark energy
models, large variations in the dark energy density or spatial curvature
with $\ok<0$ can weaken the link between $f(z)$ and $\om(z)$ (or $H(z)$). An
observed deviation from the expected value and near constancy of $\gamma$
would certainly falsify \lcdm\ and some simple dark energy models,
but using the growth index as a test of smooth dark energy in
general may require refinement of the standard parametrization of
Eq.~(\ref{eq:gammadef}) to account for significant $w(z)$ variation and crossing
of $\om(z)=1$.  Alternatively, one can adopt the more general approach of
examining the integrated growth function at various redshifts as discussed above.

There are also other, more fundamental but more qualitative means of testing
smooth dark energy.  By definition, on scales where the dark energy remains
smooth there is no particular scale for growth in the linear regime.  Models of
acceleration that involve coupling of dark energy to dark matter or
modifications of gravity that introduce new scales in addition to the Hubble scale
generically imply scale dependent linear growth (e.g.\ \cite{WhiKoc01}).
Such models can also feature differences in dynamical and lensing
mass measurements.  Generically, modified gravity models that satisfy local
constraints on gravity also break the relationship between the linear and
nonlinear growth of structure again by the introduction of a new scale to the
problem \cite{Vai72,HuSaw07,OyaLimHu08}.

Another way in which the standard cosmological paradigm might be falsified is
through observed violations of the relation $\dlum(z)=(1+z) D(z)$ between
luminosity distances and comoving angular diameter distances.  Examples of
mechanisms for violating this relation include photon-axion mixing, photon
decay, and nonzero torsion in the gravity theory (e.g.\ \cite{Bassett_Kunz_03,
  Uzan04,SonHu05}). Therefore, the ``duality relation'' between the two
distances is an interesting test of exotic new physics possibly related to
acceleration. There also exists a more general (but related) consistency
relation between the comoving distance $D(z)$ and the Hubble parameter $H(z)$
that holds in any homogeneous and isotropic FRW model \cite{Clarkson_Bassett}
and can be tested using accurate cosmological observations of the two
functions at any redshift.  Our standard assumptions could also be falsified
through observed violations of homogeneity or isotropy signaling a breakdown
in the validity of the FRW metric, or by observing time variation in
fundamental constants.

We have shown that dark energy degrees of freedom and spatial
curvature permit the basic distance, expansion, and growth observables to vary
greatly. Nevertheless, 
there are still many ways in which the dark energy paradigm for
acceleration could be falsified.

%=================================================================
\section{Conclusions}
\label{sec:conc}

% ****************************************
\begin{figure*}[t]
\centerline{\psfig{file=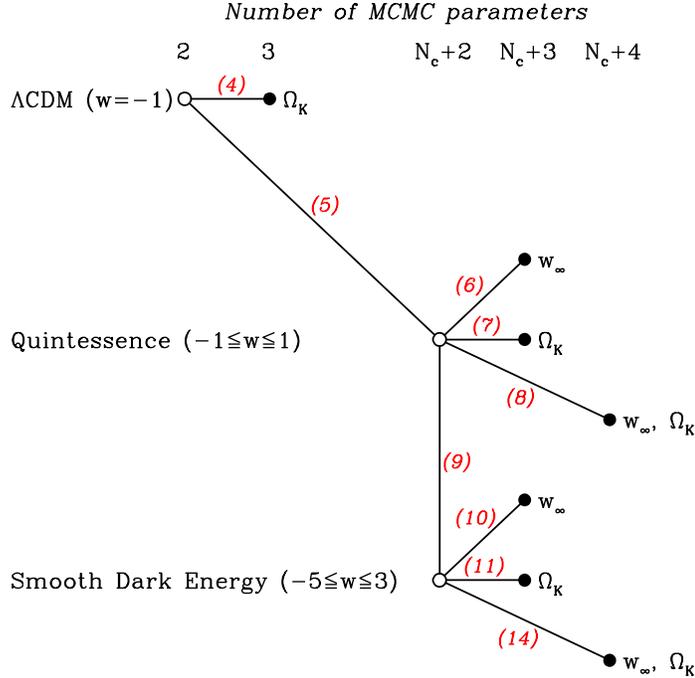, width=4.in}}
\caption{Index for dark energy model classes compared in 
figures in \S~\ref{sec:dectree}. 
The number of parameters 
varied in the MCMC likelihood analysis increases 
from left to right. For extensions to the 
baseline model within each class, the additional parameters ($\ok$, 
$\winf$, or both) are listed. Red numbers along lines connecting 
two models indicate the number of the figure in this paper in which 
we plot growth and expansion predictions for that pair of models.
}
\vskip 0.25cm
\label{fig:index}
\end{figure*}
% ****************************************

Using a combination of quantitative tools including principal components of the
dark energy equation of state and MCMC analysis with 
simulated future data, we demonstrate that combined constraints on dark energy from the
measurement of distances, growth, and the expansion rate provide many ways to test 
not only specific dark energy models, but also general \emph{classes} of models.
In particular, a high-quality supernova sample such as
that anticipated from SNAP and the CMB data expected from Planck make strong
predictions for other observables in the context of a wide variety of models.
Follow-up observations of the predicted observables offer the exciting
possibility of falsifying model classes and demonstrating the need
for a new paradigm for acceleration.

Figure~\ref{fig:index} summarizes the basic dark energy model classes and their 
generalizations, arranged by the allowed
values of $w$ and the total number of MCMC parameters 
[Eq.~(\ref{eq:parameters})]. 
In general, observations that falsify classes toward the upper left
corner of this ``tree'' of model classes 
require adding more freedom to the models by moving down and/or 
to the right in the tree.
This diagram serves as an index to the figures in \S~\ref{sec:dectree}; 
each line between a pair of model classes is labeled by the number of
the figure in which a comparison of observable predictions for those two
classes may be found.

In the context of the current standard model, flat \lcdm, predictions for
acceleration observables come mainly from the CMB. The SNAP SN data themselves
provide one stringent test of flat \lcdm\ predictions from Planck.  Other
types of observations are also well suited for testing flat \lcdm.  In
particular, constraints from future SN and CMB data have a 
narrow pivot point in the
expansion rate where $H(z=1)$ is predicted to $\sim 0.1\%$ (at 68\% CL),
making such a measurement an especially interesting
target for future BAO experiments and other probes of $H$.  Likewise, growth
observables are predicted to better than $0.5\%$ at all redshifts.

Even if we drop the assumption of flatness, \lcdm\ remains highly predictive.
The pivot point in $H(z)$ predictions disappears
when nonzero spatial curvature is allowed, but the \lcdm\ model class can
still be falsified with $\gtrsim 1\%$ deviations in any of the acceleration
observables at any redshift.

Remarkably, allowing general time variation of the dark energy equation of
state $w(z)$ within the class of quintessence models ($-1\leq w\leq 1$) does
not significantly weaken predictions if we assume flatness.
For these models, SN
and CMB data predict growth and absolute distance to $\sim 1-2\%$ precision.

Including the possibility of early dark energy, e.g.\ a scalar field that
tracks the matter density at high redshift, does not weaken these predictions
substantially as long as the SN and CMB observations remain consistent with
a cosmological constant as current data suggest. 
An increased fraction of early dark energy reduces
the distance to recombination, $\dlss$, and must be compensated by either
allowing nonzero curvature or reducing the dark energy density at late
times. The latter option requires $w<-1$ which is forbidden for quintessence
models with positive potentials, so a large fraction of early dark energy is
not allowed for quintessence in a flat universe.

The same $w=-1$ barrier for quintessence leads to predictions of one-sided
deviations from flat \lcdm\ observables when spatial curvature is allowed to
vary. Closed universes
reduce $\dlss$, so matching CMB constraints requires the ability to lower the
dark energy density with $w<-1$. Therefore, quintessence predictions from SN
and CMB data consistent with $w=-1$ favor \emph{open} universes.  
Curvature in open universes causes additional growth suppression, 
particularly at $z \lesssim 2$.

Open quintessence models can also have significant early dark energy that
suppresses growth by a constant factor at low redshift.
Even with the additional freedom in curvature and
early dark energy, SN and CMB data still provide general predictions for
quintessence. For example, neither the growth relative to recombination nor
absolute distances at low redshift can be significantly larger than in flat
\lcdm.  Growth relative to the present out to $z\sim 2$ cannot differ 
from its standard behavior by more than $\sim 2-4\%$.

Allowing the low-redshift dark energy equation of state to vary beyond the
range of quintessence enables new types of models since the absence of the
$w\geq -1$ bound permits deviations in dark energy density both above and below
the constant density of flat \lcdm. These general smooth dark energy models are therefore able
to have significant early dark energy in a flat universe,
and closed universes consistent with the SN and CMB
data are also possible. As a result, smooth dark
energy models can have both larger and smaller growth and absolute distances
relative to flat \lcdm, unlike the one-sided predictions of quintessence.

Even the most general model class including large dark energy variations at low
redshift, early dark energy, {\it and} nonzero curvature makes some generic
predictions given future SN and CMB data.  Growth at
$z\lesssim 2$ normalized to its present value can be no more than $\sim 5\%$
larger and the expansion rate at $z\gtrsim 2$ no more than $\sim 5\%$ smaller
than their values in a flat \lcdm\ cosmology.  The growth index $\gamma$ as it
is typically defined is tightly constrained by SN and CMB data in \lcdm,
but not for more general dark energy models since both time variation of 
the dark energy equation of state and the possibility of crossing
$\om(z)=1$ in closed universes weaken the relation between growth rate and
matter fraction.  Fortunately, growth measurements at different redshifts
still give tight predictions since the growth evolution at low redshift 
depends mainly on curvature.  Additional tests outside the scope
of this work, such as searching for scale dependence of linear growth, 
could also falsify all smooth dark energy models.  Falsification of the
most general smooth dark energy predictions would require new 
paradigms for cosmic acceleration and possibly
even gravity itself.

{\it Acknowledgments:}
We thank Eric Linder, Hiranya Peiris, and Amol Upadhye for useful conversations.
MM and WH were supported by the David
and Lucile Packard Foundation and the KICP under NSF PHY-0114422.
MM was additionally supported by an NSF Graduate Research Fellowship. 
WH was additionally supported by DOE contract DE-FG02-90ER-40560. 
DH was supported by the DOE OJI grant under contract DE-FG02-95ER40899, and
NSF under contract AST-0807564.

\appendix

%=================================================================
\section{Fisher Matrices and Likelihood Functions}
\label{app:data}

In this Appendix, we describe the (future) cosmological data and priors that we assume
for this study. We give expressions for the Fisher matrices for these data, 
which we use to compute the principal components of the dark energy 
equation of state, and for the likelihood functions that we use in  
MCMC analysis.

The Fisher matrix for supernovae is \cite{TegEisHuKron}
\begin{equation}
F_{ij}^{\rm SN} = \sum_{\alpha} \sigma_{\alpha}^{-2} 
\frac{dm(z_{\alpha})}{d\theta_i}\frac{dm(z_{\alpha})}{d\theta_j},
\end{equation}
where $m(z_{\alpha})= 5\log[H_0 \dlum(z_\alpha)] + \mathcal{M}$ 
is the average magnitude of the SNe 
in the redshift bin denoted by $z_{\alpha}$, $\sigma_{\alpha}$ is the error in the average magnitude, and $\mathcal{M}=M-5\log(H_0/{\rm Mpc}^{-1})+25$ 
is a constant related to the unknown absolute magnitude of the SNe.

For the fiducial supernova data, we take the expected redshift distribution
for SNAP \cite{KLMM_SNAP} plus a low-$z$ sample of 300 SNe at $0.03<z<0.1$.
The SNAP magnitude errors include both statistical and systematic components:
\begin{equation}
\sigma_\alpha^2 = \left(\frac{\dz}{\dzsub}\right)\left[\frac{0.15^2}{N_\alpha} +
0.02^2 \left(\frac{1+z}{2.7}\right)^2 \right],
\label{eq:magerr}
\end{equation}
where $N_\alpha$ is the number of SNe in each bin of width $\dz$ 
($\dz=0.1$ except for the statistical uncertainties in 
the low-$z$ SN bin, for which $\dz=0.1-\zminsn=0.07$), 
and $\dzsub$ is the width of the sub-bins used to smooth the distribution 
of SNe in redshift. We use 500 sub-bins up to $\zmax=1.7$. 
The second term on the right hand side of Eq.~(\ref{eq:magerr}) models
a systematic floor that increases linearly with $z$ up to
a maximum at $\zmax$ of $0.02$ mag per $\dz=0.1$ bin \cite{LinHut_highz}.

For the Planck CMB constraint, we start with the $2\times 2$ covariance matrix
$\tilde{C}^{\rm CMB}$ for the parameters 
\begin{equation}
\bm{\tilde{\theta}} = \{\ln
(\dlss/{\rm Mpc}),\omhh \}\,.
\label{eq:tildetheta}
\end{equation}
Here $\dlss$ is the comoving angular diameter distance to recombination.  We
ignore additional CMB information about dark energy from the ISW effect except
in the current prior on early dark energy described below.  The elements of
the covariance matrix are $\tilde{C}^{\rm CMB}_{11}=(0.0018)^2$ and
$\tilde{C}^{\rm CMB}_{22}=(0.0011)^2$ and $\tilde{C}^{\rm
  CMB}_{12}=-(0.0014)^2$.  Rotating to the space of MCMC parameters, e.g.
$\bm{\theta}_{\rm full}$ [Eq.~(\ref{eq:parameters})] gives $F^{\rm CMB} = D
[\tilde{C}^{\rm CMB} ]^{-1} D^{\rm T}$, where $D_{ij} =
d\tilde{\theta}_j/d\theta_i$. As we shall see, for the likelihood evaluation
it is more convenient to project the MCMC parameters onto the original basis
of Eq.~(\ref{eq:tildetheta}).  For a similar treatment of CMB priors on dark
energy models, see Refs.~\cite{Hu_standards,Dick}.

Priors on additional parameters can be included by adding the assumed inverse 
covariance to the appropriate entry of the Fisher matrix. 
The Fisher matrix for the full set of parameters is 
\begin{equation}
F_{ij}^{\rm tot} = F_{ij}^{\rm SN}+F_{ij}^{\rm CMB}+F_{ij}^{\rm prior}.
\label{eq:fisher}
\end{equation}
The priors and the parameters in the Fisher matrix
depend on the particular application and differ between
the PC construction and the the likelihood analysis.
The procedure for computing PCs from the Fisher 
matrices and the assumed priors are described in Appendix~\ref{app:pcs}.

For the MCMC analysis described in \S~\ref{sec:mcmc} we
 assume a Gaussian likelihood,
${\mathcal L} \propto \exp(-\chi^2/2)$, described by
\begin{equation}
\chi^2 = \chi^2_{\rm SN} + \chi^2_{\rm CMB} + \chi^2_{\rm prior}
\label{eq:chisq_tot}
\end{equation}
which includes contributions from the SNAP SN data, Planck CMB data, and
our external priors.  

We model the SN $\chi^2$ term as
\begin{eqnarray}
\chi^2_{\rm SN} =&& A-\frac{B^2}{C},\label{eq:chisqsn}\\
&& A = 5 \sum_{\alpha} \frac{[\Delta \log(H_0 \dlum(z_{\alpha}))]^2}
{\sigma_{\alpha}^2},\nonumber\\
&& B = 5 \sum_{\alpha} \frac{\Delta\log(H_0 \dlum(z_{\alpha}))}
{\sigma_{\alpha}^2}, \nonumber \\
&& C = \sum_{\alpha} \frac{1}{\sigma_{\alpha}^2}, \nonumber
\end{eqnarray}
where $\Delta \log (H_0 \dlum)$ refers to the difference between a model 
$H_0\dlum$ derived from MCMC parameters and the fiducial value.
 The variance $\sigma_{\alpha}^2$ is modeled as in Eq.~(\ref{eq:magerr}).

 The $B^2/C$ term in $\chi^2_{\rm SN}$ comes from the marginalization over
 $\mathcal{M}$.  Because of this marginalization, the SN data are insensitive
 to redshift-independent shifts in the magnitudes $m(z_{\alpha})$ caused by changes in
 combinations of other cosmological parameters, e.g.\ $\om$ and the PC
 amplitudes for $w(z)$. This shift in magnitudes corresponds to multiplying
 the distances by a constant factor: $H_0\dlum(z) \to (1+\delta)H_0\dlum(z)$.

The CMB contribution to $\chi^2$ is
\begin{equation}
\chi^2_{\rm CMB} = \sum_{i,j=1}^2 \delta \tilde\theta_i [C^{\rm CMB}_{ij}]^{-1}  \delta \tilde\theta_j, 
\label{eq:chisqcmb} 
\end{equation}
where $\tilde {\bf \theta}_i$ is the same as in Eq.~(\ref{eq:tildetheta}) 
and $\delta {\tilde{\theta}_i} ={\tilde{\theta}_i} - {\tilde{\theta}_i}|_{\rm fid}$ is the
difference in $\ln \dlss$ and $\omhh$ from the fiducial model 
with $\dlss$ derived from the MCMC parameters.

When computing $\chi^2$ we set the SN magnitudes, $\dlss$, and $\omhh$ to have
the exact values predicted for the fiducial cosmology without any scatter in
the simulated measurements. The resulting constraints from the data are
therefore expected to be centered on the fiducial parameter values rather than
shifted away from them by $\sim 1~\sigma$. We do this because we are mainly
interested in the width of predictions for observables and not their central
values for a particular realization of the measurement errors.

In the MCMC likelihood analysis we employ three external priors based on {\it current} data
to limit our study to reasonable cosmologies, and one internal theoretical prior:
\begin{equation}
\chi^2_{\rm prior} = \chi^2_{\rm H} + \chi^2_{\rm BAO} + 
\chi^2_{\rm EDE} +\chi^2_{w},
\label{eq:chisqprior}
\end{equation}
where the terms on the right hand side respectively refer to an
HST Key Project prior on the Hubble constant of width $\sigma(h) = 0.08$ 
\cite{HKP}, 
a BAO prior on the angular diameter distance to $z=0.35$ with 
$\sigma(\ln D(z=0.35))=0.037$, roughly corresponding to the 
constraint from the SDSS LRG sample \cite{Eisenstein}, and 
 a WMAP prior on the fraction of early 
dark energy with $\sigma(\ode(z_*))=0.025$, based on the constraints 
on early dark energy models in Ref.~\cite{Doran}.

In addition to constraints from current data, the last term in 
Eq.~(\ref{eq:chisqprior}) includes 
theoretical limits on the dark energy equation of state, 
$\wmin\leq w\leq \wmax$. These limits are typically implemented 
as an infinite barrier in $\chi^2_w$ corresponding to some range
allowed by the model class, i.e. a top hat prior on the 
MCMC parameters. 
To compute these priors, we start with the projection of 
$w(z)$ onto PC  amplitudes,
\begin{equation}
\alpha_i = \frac{1}{\nzpc}\sum_{j=1}^{\nzpc}[w(z_j)-\wfid]e_i(z_j),
\end{equation}
where $\wfid=-1$ unless otherwise specified. 
By finding the values of $w(z_j)$ within the allowed range $[\wmin,\wmax]$ 
that maximize or minimize $\alpha_i$, we obtain limits on 
the amplitude of each PC, $\alpha_i^{(-)} \leq \alpha_i \leq \alpha_i^{(+)}$,
where
\begin{eqnarray}
&&\alpha_i^{(\pm)} \equiv \frac{1}{2\nzpc} \sum_{j=1}^{\nzpc}
[(\wmin+\wmax-2\wfid)e_i(z_j) \nonumber\\
&&\qquad \pm (\wmax-\wmin) |e_i(z_j)|\,].\label{eq:prior1}
\end{eqnarray}
The width of this prior, $\alpha_i^{(+)}-\alpha_i^{(-)}$, depends  
on the width of the allowed range of $w$ but not the value of $\wfid$.

% ****************************************
\begin{figure}[t]
\centerline{\psfig{file=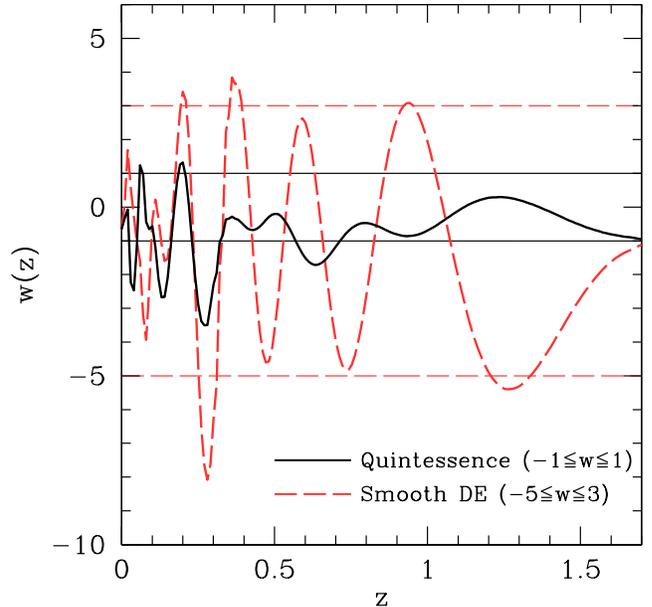, width=3.5in}}
\caption{Examples of equations of state constructed from 
the 15 PCs in Fig.~(\ref{fig:pcs}), where the PC amplitudes 
satisfy the priors given by 
Eqs.~(\ref{eq:prior1}) and~(\ref{eq:prior2}) for $-1\leq w\leq 1$ 
(\emph{solid black}) and $-5\leq w\leq 3$ (\emph{dashed red}).
}
\vskip 0.25cm
\label{fig:wzpriors}
\end{figure}
% ****************************************

We find a second prior on $\{\alpha_i\}$ 
using the fact that restricting the range of $w(z)$
places an upper limit on $\sum_i [w(z_i)-\wfid]^2$. From Eqs.~(\ref{eq:pcstow}) 
and~(\ref{eq:norm}), 
\begin{equation}
\sum_{i=1}^{\nzpc} [w(z_i)-\wfid]^2 = \nzpc \sum_{i=1}^{\nzpc} \alpha_i^2.
\label{eq:prior2a}
\end{equation}
The bounds on $w(z)$ impose an upper limit on this sum,
\begin{eqnarray}
[w(z_i)-\wfid]^2
& \leq & V_{\rm max},\label{eq:prior2b}\\
 V_{\rm max} &\equiv&
 {\rm max} [(\wmax-\wfid)^2, (\wmin-\wfid)^2]. \nonumber
   \end{eqnarray}
Combining Eqs.~(\ref{eq:prior2a}) and~(\ref{eq:prior2b}) we find that 
the PC amplitudes must lie within a sphere in the parameter space:
\begin{equation}
\label{eq:prior2}
\sum_{i=1}^{N_c} \alpha_i^2 \leq
\sum_{i=1}^{\nzpc} \alpha_i^2 
 \leq V_{\rm max},
\end{equation}
where $N_c<\nzpc$ is the number of components in the truncated set 
of PCs used for likelihood analysis (see \S~\ref{sec:pcs}).
For a given allowed range of $w$, the constraint on PC amplitudes 
from this inequality is strongest when the fiducial model lies 
in the center of that range, $\wfid = (\wmin+\wmax)/2$, but 
in general the limits in Eq.~(\ref{eq:prior1}) are the stronger 
of the two PC priors. 

Figure~\ref{fig:wzpriors} shows examples of $w(z)$ parametrized by 15 PCs 
that satisfy the bounds in Eqs.~(\ref{eq:prior1}) and~(\ref{eq:prior2})
corresponding to the quintessence model class, with $\wmin=-1$ and 
$\wmax=1$, and the smooth dark energy class, with $\wmin=-5$ and 
$\wmax=3$. In each case, the sum of the 15 components is 
allowed to violate the bounds on $w$ at some redshifts. For some 
models, the addition of higher-variance PCs can correct for these 
deviations from the allowed range so that $\wmin\leq w\leq \wmax$ 
everywhere. However, this is not necessarily true for all 
models that satisfy the PC priors; the most important property of these 
priors is that they retain any model that {\it does} obey the limits on $w$ 
while excluding a large number of unacceptable models.
For more details on the derivation of these priors, see Ref.~\cite{MorHu08} 
where similar bounds are used in the context of reionization models 
to constrain the ionized fraction of hydrogen to the range $0\leq x_e\leq 1$.

When switching to alternate priors to test volume effects related 
to curvature as described in \S~\ref{sec:tree3}, we transform the 
PC priors either by adding an additional term to the likelihood 
in the MCMC analysis or by modifying the weights of samples in the 
chain as a post-processing step. In either case we convert the 
usual top-hat PC priors to ones that are flat in the density of each 
PC at $\zmax$ relative to $z=0$ [Eq.~\ref{eq:flatrhoprior}] by multiplying 
$\mathcal{L}$ or the sample weight by $\exp(-\chi^2_{\rm PC}/2)$, where
\begin{equation}
\chi^2_{\rm PC} = -2 \sum_i \left[3\alpha_i \int_0^{\zmax} dz \frac{e_i(z)}{1+z}\right].
\end{equation}

In addition to priors on the low-$z$ equation of state, we place 
a top-hat prior on the early dark energy at $z>\zmax$ 
corresponding to $\wmin\leq \winf\leq \min(0,\wmax)$.  Since the early dark 
energy MCMC parameter is actually $e^{\winf}$, in practice the 
prior we use for MCMC is $e^{-1}\leq e^{\winf} \leq 0$ for quintessence 
and $e^{-5}\leq e^{\winf} \leq 0$ for smooth dark energy.

%=================================================================
\section{Principal component methodology}
\label{app:pcs}

To compute the principal components of $w(z)$ from the Fisher matrices of
Appendix~\ref{app:data} (with $F_{ij}^{\rm prior}$ specified below), we first
take the $\nzpc\times \nzpc$ submatrix of $(F_{ij}^{\rm tot})^{-1}$
corresponding to the redshift binned dark energy equation of state, $w(z_i)$.
We then invert the $\nzpc\times \nzpc$ matrix to get $F^w_{ij}$, which is the
original Fisher matrix marginalized over everything except $\{w(z_i)\}$.
Finally, we compute the eigenvectors of $F^w_{ij}$, which are the PC functions
and normalized as in Eq.~(\ref{eq:norm}), and the eigenvalues, which are the
inverse variances of the PC amplitudes.

We evaluate $F_{ij}^{\rm tot}$ at the same fiducial model as for the MCMC likelihood,
i.e. flat \lcdm\ with $\om=0.24$ and $h=0.73$. This model is consistent with current data, 
which should minimize the number of principal components needed to accurately parametrize 
viable dark energy models.  The exact choice of fiducial cosmology is unimportant as 
principal component shapes do not vary greatly with changes in the fiducial model 
that are consistent with current data.

Our default binning scheme is $\nzpc=500$ bins between $z=0$ and $\zmax$. 
These bins are fine enough to obtain reasonably continuous PC shapes and 
to allow varying $w(z)$ at $z<\zminsn$, which has important consequences 
as we describe later.

We choose $\zmax=1.7$ to match the assumed maximum redshift of the SN sample. 
A smaller choice of $\zmax$ would not significantly change the PC shapes at lower $z$ 
but would result in a less complete set of PCs due to neglecting some of the SN data.
Increasing $\zmax$ would require additional support from SNe or other data at 
$z>1.7$ for the PCs to have any weight at higher redshift; the CMB distance 
constraint helps somewhat but is still only a single data point for constraining 
the additional $w(z_i)$. Furthermore, if the
expansion history at low $z$ is near the fiducial flat \lcdm\ model then the
lack of weight in the PCs at high $z$ is mainly a consequence of dark energy becoming less
significant as redshift increases \cite{Huterer_Starkman}.

Unlike the MCMC likelihood analysis, we do not include external priors 
from current data
or priors on PC amplitudes in the $F_{ij}^{\rm prior}$ term.
However, we do use priors that correspond to fixing certain parameters 
besides $\{w(z_i)\}$, i.e. $\om$, $\ok$, and $\winf$. 
In the rest of this section we explain our choices 
of which of the other parameters are fixed and which are marginalized over.
Note that we do not consider $\omhh$ here because it is nearly fixed 
automatically due to the constraint in $F_{ij}^{\rm CMB}$.

The question of fixing or marginalizing parameters is essentially a question 
of whether or not we wish to include degeneracies between those parameters 
and $w(z)$ in the low-variance PCs that we retain for MCMC analysis.
Marginalizing these other parameters eliminates from the low-variance 
PCs the modes in $w(z)$
that have a degenerate effect in the SN distances and CMB data but are not 
necessarily degenerate in other acceleration observables.  
Since this marginalization generally results in an incomplete PC
basis, the better choice is typically to fix the non-$w(z)$ parameters 
with $F_{ij}^{\rm prior}$ so that the modes of $w(z)$ degenerate with them 
are assumed to be well measured and therefore are included in the low-variance PCs.
Fixing these additional parameters ensures that the PCs we use are as complete as 
possible in the acceleration observables, and so we fix 
$\ok$ and $\winf$ when computing the PCs. However, $\om$ is an exception to this 
rule where having a fully complete basis for $w(z)$ is not desirable.

There are two types of degeneracy that can in principle exist between $\om$ and $w(z)$ 
in the assumed SN and CMB data.  First,  the  dark energy can mimic 
some fraction of the matter density (leaving $H_0\dlum(z)$ unchanged) 
by approaching $w=0$ at high $z$ (e.g.\ \cite{Kunz07}). 
In the context of a spatially flat geometry
with no early dark energy, this degeneracy is eliminated by the CMB constraints on $\dlss$ 
and $\omhh$.   The impact of these constraints weakens if we 
allow for the freedom to adjust either curvature or early dark energy.  
By constructing PCs with curvature and early dark energy fixed, we 
assume that the matter-mimicking mode of $w(z)$ can be measured by the SN distances
and CMB constraints. This assumption ensures that the PCs are complete 
with respect to this mode, regardless of whether we marginalize or fix $\om$ in 
the PC construction.
This type of degeneracy between $\om$ and $w(z)$ is therefore properly included in 
the MCMC predictions for model classes with curvature and early dark energy.

A second type of degeneracy is introduced by the minimum redshift for which SNe can
be measured in the Hubble flow.  The ability to determine $H_0\dlum(z)$ from SN observations 
depends on how well we can anchor the relative distances to $z=0$. 
As noted in Appendix~\ref{app:data}, marginalization over the nuisance parameter 
$\scrm$ causes the SN observations to be insensitive to constant shifts in relative 
distances of the form
\begin{eqnarray}
H_0\dlum(z)&\to& (1+\delta)H_0\dlum(z), \label{eq:snshift}\\
\scrm&\to&\scrm-5\log(1+\delta).\nonumber
\end{eqnarray}
We are not free to change $H_0\dlum(z)$ at all redshifts 
since $H_0\dlum(z)=z$ at low $z$ independent of the cosmology.
However, it is possible (if unlikely) 
that a large variation in $w(z)$ near $z=0$ changes the SN distances 
by a nearly constant factor at all redshifts 
except at $z \lesssim \zminsn$, where $\zminsn$ is the minimum SN redshift.
When including the CMB data this effect creates a degeneracy between $w(z)$ and $\om$ 
since the shift in Eq.~(\ref{eq:snshift}) requires changes in $\om$ and $H_0$ to satisfy 
CMB constraints on $\dlss$ and $\omhh$.

If we fix $\om$ when computing the PCs, then 
we assume that the behavior of $w(z)$ at $z<\zminsn$ is well constrained and therefore 
large variations in the equation of state at low redshift are included in the 
low-variance PCs. 
Instead, we choose to reduce the impact of this degenerate mode of $w(z)$ by 
marginalizing $\om$ in the PC construction. Our basis for $w(z)$ is therefore 
incomplete with respect to this mode, but there are several reasons for neglecting 
large variations in $w(z)$ at $z<\zminsn$.
One benefit of this approach is that by reducing the degeneracy 
between $\om$ and the PC amplitudes we improve convergence of MCMC samples; in 
the presence of the full degeneracy, it is difficult to obtain well-converged chains
even for the simplest class of models with PC-parametrized $w(z)$.
Furthermore, apart from the Hubble constant itself and the interpretation of 
SN data as measuring $H_0 \dlum(z)$ as opposed to $\dlum(z)/\dlum(\zminsn)$, 
acceleration observables are not significantly affected by this degeneracy. 
Finally, despite marginalizing $\om$ when computing PCs we still 
retain enough of this degeneracy that predictions for the $z\rightarrow 0$ 
behavior of $D(z)$ and $H(z)$ is appropriately uncertain and limited by our prior on $H_0$, 
as shown in Fig.~\ref{fig:zdist1_1}, for example.

Improving measurements of $H_0$
beyond the current level would further limit the possibility of these ultra-low redshift
changes in $w$.  Conversely, in the absence of such variation in $w(z)$, 
precision $H_0$ measurements
play the same role as low redshift $D$ and $H$ measurements in all cases considered
in the main paper. 

In summary, when constructing the PCs we take priors that
 fix $\ok=0$ and $\winf=-1$ but marginalize $\om$.  We do not employ the additional 
 current priors from BAO distance to $z=0.35$, HST Key Project measurement of $H_0$, 
or WMAP limits on early dark energy that are added to the likelihood analysis.

%=================================================================
\section{Completeness Tests}
\label{app:complete}

When making predictions for general classes of models, we need to make sure that 
the parametrization we use has sufficient freedom to explore all types of effects 
that models can have on the acceleration observables. 
In this appendix we present several tests of the completeness of our parametrization.
We begin by justifying the number of principal components of $w(z)$ used in 
the MCMC likelihood analysis. We then examine the sensitivity of our results 
to the choices of fiducial model and maximum redshift for principal components.
Finally, we discuss the limitations of our early dark energy parametrization.

The PCs form a complete basis for $w(z)$ ordered by how well 
they can be measured by  the fiducial SN and CMB data. 
This ordering allows us to truncate the set of PCs to some small number 
that have the greatest impact on the fiducial data. Retaining a limited  
number of PCs is a practical necessity to make parameter estimation 
feasible, but we must make sure that the higher-variance PCs that we 
ignore do not make significant contributions to the expansion or growth observables
predicted in the main paper.

It is important to emphasize that we do not expect or demand completeness in
unobservable quantities like $w(z)$ itself; the high-variance PCs that we
neglect can have large effects on the equation of state, but since all of the
observables contain integrals over $w(z)$ the effects of these rapidly
oscillating PCs (see Fig.~\ref{fig:pcs}) tend to cancel out for the
redshift-dependent quantities of interest. This is especially true for 
distances and integrated growth, each of which involves essentially two
integrals of $w(z)$ over redshift.  Completeness is more difficult to 
attain for observables with a single redshift integral such as the expansion rate 
and growth rate, but the practical requirement of a large volume
for such observations makes any rapid evolution with redshift unobservable in
practice.

% ****************************************
\begin{figure}[t]
\centerline{\psfig{file=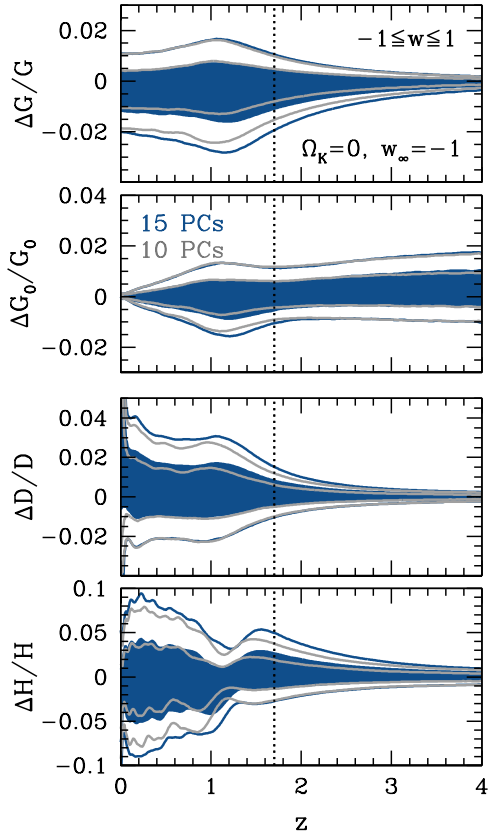, width=2.7in}}
\caption{Test of PC completeness for quintessence models. 
Predictions for growth and expansion observables from 
MCMC with 10 (\emph{light gray}) and 15 (\emph{dark blue}) PCs
for flat quintessence models ($-1\leq w \leq 1$) with 
$\winf=-1$ (no early dark energy).
}
\vskip 0.25cm
\label{fig:comp1}
\end{figure}
% ****************************************

% ****************************************
\begin{figure}[t]
\centerline{\psfig{file=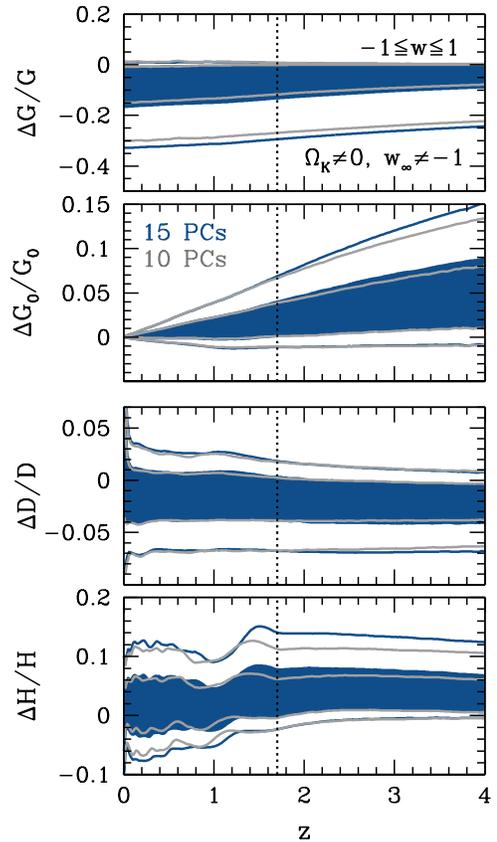, width=2.7in}}
\caption{Same as Fig.~\ref{fig:comp1} for non-flat quintessence 
models with early dark energy.
}
\vskip 0.25cm
\label{fig:comp2}
\end{figure}
% ****************************************

% ****************************************
\begin{figure}[t]
\centerline{\psfig{file=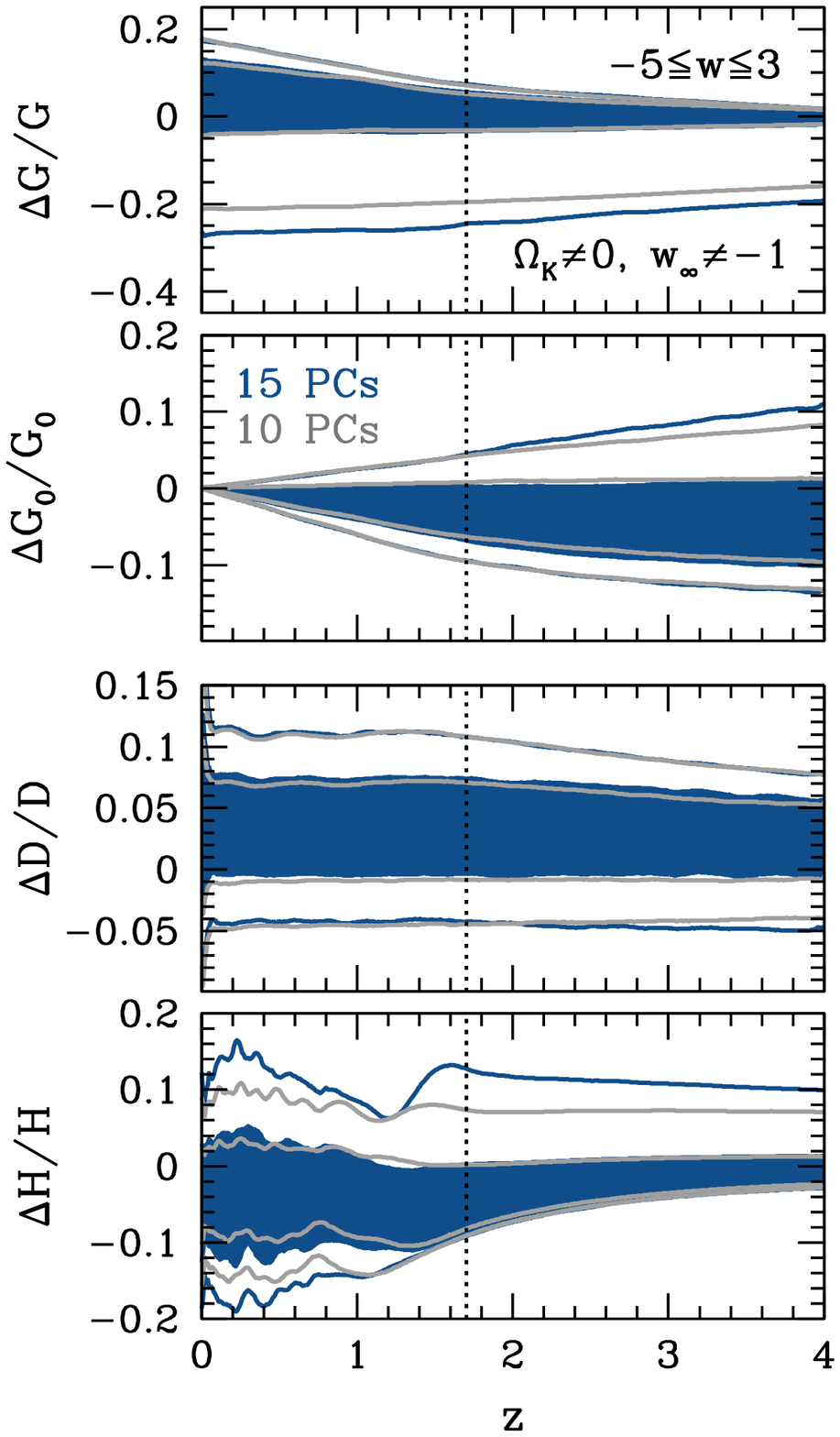, width=2.7in}}
\caption{Same as Fig.~\ref{fig:comp1} for smooth dark energy models
($-5\leq w\leq 3$) including curvature and early dark energy.
}
\vskip 0.25cm
\label{fig:comp3}
\end{figure}
% ****************************************

Our basic strategy for determining the number of PCs required for 
completeness, $N_c$, is to repeat the MCMC analysis for each class of models 
using varying numbers of PCs. As we increase the number of components 
of $w(z)$, we expect the resulting predictions for observables to 
eventually converge once we have reached the necessary number of PCs. 
This approach makes the value of $N_c$ to some extent dependent on 
what we assume about the data and the allowed models. For example, it 
may be that the $N_c+1$ component can have a significant effect on 
certain observables if its amplitude $\alpha_{N_c+1}$ is unconstrained, 
but limits on this amplitude from the data and/or priors keep the 
impact of this eigenmode on the observables small. Similarly, it is 
possible that the value of $N_c$ changes as we add the additional 
freedom of spatial curvature and early dark energy to the baseline model.

Since our definition of completeness is based on the precision of predictions
for a variety of acceleration observables, $N_c$ will generally differ from
(and be larger than) the number of dark energy parameters that can be measured
to some specified accuracy
\cite{Linder_Huterer_howmany,Albrecht_Bernstein,Suletal07} or the number of
parameters required by the data in a Bayesian model selection sense
\cite[e.g.,][]{Liddle_evidence,Davetal07}.

Figure~\ref{fig:comp1} shows a comparison of observable predictions 
for the baseline quintessence class  ($\ok=0$, $\winf=-1$), using 
either 10 or 15 PCs. There is little difference between the two sets 
of predictions, suggesting that $N_c\sim 10$ is sufficient for 
this model class. In contrast, predictions in this class using only 
5 PCs are significantly tighter than those with 10 PCs.

Including both curvature and early dark energy in the quintessence model class
does not alter the agreement between the predictions for 10 and 15 PCs, 
as shown in Fig.~\ref{fig:comp2}.
There is slightly more variation in the limits on observables at $z>\zmax$, 
but we do not expect perfect completeness at high redshift anyway due to 
our simplistic early dark energy parametrization (see below).
Even in our most general model class where we weaken the quintessence prior to 
$-5\leq w\leq 3$ while continuing to include curvature and early dark energy, 
as in Fig.~\ref{fig:comp3}, the predictions remain robust to increasing the 
number of PCs from 10 to 15.

These comparisons indicate that $N_c \sim 10$ is sufficient for completeness in 
all model classes that we study here. In the main sections of this paper we  
present results from the larger, ``overcomplete'' set of 15 PCs.

In our predictions throughout this paper, we have assumed a particular flat \lcdm\ 
model both for PC construction and for creating the fiducial SN and CMB data for 
MCMC likelihoods. Since the true cosmology could be somewhat different, we can ask 
how the predictions for observables would change had we assumed a different 
fiducial model. To test this dependence, we have redone the MCMC analysis 
using fiducial cosmologies with various values of constant $\wfid \ne -1$. 
For simulated SN and CMB data based on $\wfid = -0.93$ (approximately the 68\% upper limit 
of combined constraints on constant $w$ from WMAP and current BAO and SN data 
\cite{Komatsu_2008}), the predictions for growth and expansion observables 
in the context of quintessence models with either curvature or early dark energy 
have similar uncertainties to those with a fiducial \lcdm\ cosmology (Figs.~\ref{fig:zdist1_2}
and~\ref{fig:zdist1_3}). The main effect of increasing $\wfid$ is that it slightly 
weakens the impact of the quintessence $w=-1$ barrier; moving $\wfid$ away from this 
barrier makes it possible to slightly reduce the dark energy density at $z<\zmax$ 
from its fiducial level, and therefore the predictions for 
quintessence models include some features that were previously only allowed in the 
more general smooth dark energy class. For example, compared with the flat \lcdm\ 
growth history, flat quintessence models with 
early dark energy are allowed to have $\Delta G/G \sim -5\%$ at 68\% CL instead of 
$-2\%$, and curved quintessence models with no early dark energy can have $\Delta G/G\sim 3\%$ 
at $z=0$ (68\% CL) where previously only downward variations in $G(z=0)$ were 
possible in this class of models. The ability to falsify certain model classes 
therefore depends on how consistent future SN and CMB data sets are with the standard 
flat \lcdm\ cosmology; any significant variation would be interesting in its own right
and would make some changes to the model testing results presented here but not the
methodology or the logic of the results themselves.

Another technical issue related to completeness is whether our choice of $\zmax=1.7$ 
affects the predictions for observables. This choice enters into both the definition of 
$w(z)$ principal components as discussed in Appendix~\ref{app:pcs} and the likelihood 
for MCMC as the maximum redshift of the fiducial SN sample. To distinguish between 
the two, let us call the maximum redshift for PCs $\zmaxpc$ and for the MCMC 
likelihood $\zmaxlike$. Note that the choice of $\zmaxpc$ also influences our definition 
of ``early dark energy'' by setting the minimum redshift at which $w=\winf$.

If we keep $\zmaxlike=1.7$ but extend the PCs to $\zmaxpc=2.5$ (assuming a flat SN distribution 
at $1.7\leq z\leq 2.5$ with the number per bin equal to the number at $z=1.7$ in the 
original distribution), the resulting predictions for observables in the flat, no early 
dark energy quintessence class are similar to those in Fig.~\ref{fig:zdist1_1}. 
The predictions are slightly weaker, particularly at 95\% CL, due to the extra freedom 
in $w(z)$ at $1.7<z<2.5$. The fact that width and redshift dependence of 
constraints on observables change little with increased $\zmaxpc$ indicates that 
our results are not strongly influenced by the choice of $\zmaxpc=1.7$.
For example, the tightening of constraints on $G$, $D$, and $H$ at high $z$ 
in Fig.~\ref{fig:zdist1_1} is more a consequence of the transition from accelerated 
expansion to deceleration at $z\sim 1$ than it is of setting $\zmaxpc=1.7$.

As another test of sensitivity to $\zmax$, we set $\zmaxpc=2.5$ as before and 
also extend the SN distribution for the MCMC analysis to $\zmaxlike=2.5$ (with constant 
number per bin at $z>1.7$ as for the PCs). The resulting predictions for flat 
quintessence without early dark energy are nearly identical at $z\gtrsim 2$ to 
the ones in Fig.~\ref{fig:zdist1_1}. At lower redshifts, observables are slightly better 
constrained due to the additional SN data; the largest changes are at $z=1$, with 
new 68\% limits of $\Delta G/G \gtrsim -1\%$ and $\Delta D/D \lesssim 1\%$ (68\% CL).
Predictions for growth and expansion observables are relatively insensitive to 
increasing both $\zmaxpc$ and $\zmaxlike$.

At redshifts above our default choice of $\zmax=1.7$, we make predictions for
observables by specifying $w(z > \zmax)$ through an early dark energy ansatz
with a constant equation of state, $w=\winf$.  While $\winf$ is not a complete
parameterization of early dark energy, it does provide guidance for
predictions.  For example, even though a constant equation of state at $z>1.7$
is a poor fit to the Albrecht-Skordis model in which a quintessence scalar
field has an exponential potential modified by a quadratic polynomial
\cite{AlbSko00}, $\winf$ as an effective parameter can nevertheless
simultaneously fit the CMB distance to the required Planck precision and the
growth function at $\zmax$ to $\sim 2\%$ accuracy.  The early dark energy
parameterization acts as a diagnostic for whether the dark energy can ever
become a substantial fraction of the energy density at $z \gtrsim \zmax$ given
CMB constraints on the distance and energy densities at $z \sim z_*$.  Our
philosophy is to use an incomplete but representative parameterization that
can be used to monitor the need for early dark energy.  If a substantial
fraction of early dark energy is required by observations under this
parameterization, then more complete descriptions and more detailed
observations will be required.

\bibliographystyle{arxiv_physrev}
%\bibliography{MortHutHu08}

\def\eprinttmppp@#1arXiv:@{#1}
\providecommand{\arxivlink[1]}{\href{http://arxiv.org/abs/#1}{arXiv:#1}}
\def\eprinttmp@#1arXiv:#2 [#3]#4@{\ifthenelse{\equal{#3}{x}}{\ifthenelse{
\equal{#1}{}}{\arxivlink{\eprinttmppp@#2@}}{\arxivlink{#1}}}{\arxivlink{#2}
  [#3]}}
\providecommand{\eprintlink}[1]{\eprinttmp@#1arXiv: [x]@}
\renewcommand{\eprint}[1]{\eprintlink{#1}}
\providecommand{\eprintmod}[1][XXXX.XXXX]{\eprintlink{#1}}
\providecommand{\adsurl}[1]{\href{#1}{ADS}}
\renewcommand{\bibinfo}[2]{\ifthenelse{\equal{#1}{isbn}}{\href{http://cosmolog%
ist.info/ISBN/#2}{#2}}{#2}}

\end{document}